\documentclass[draft,onecolumn]{IEEEtran}
\usepackage{amsmath,amstext,amssymb,epsf}
\usepackage[dvips]{graphicx}
\date{}

\renewcommand{\le}{\leq}
\renewcommand{\ge}{\geq}
\renewcommand{\emptyset}{\varnothing}

\newcommand{\A}{\mathcal A}
\newcommand{\BB}{\mathcal B}

\newcommand{\C}{\mathcal C}
\newcommand{\Q}{\mathcal Q}

\newcommand{\X}{\booln}
\newcommand{\Y}{\mathcal{Y}}
\newcommand{\dmax}{n}

\newcommand{\booln}{\{0,1\}^n}

\newcommand{\K}{C}

\newcommand{\wh}[1]{\lfloor #1 \rfloor}
\newcommand{\wwh}[1]{\lceil #1 \rceil}

\newcommand{\pair}[1]{\langle #1\rangle}

\newtheorem{theorem}{\sc Theorem}

\newtheorem{lemma}{\sc Lemma}
\newtheorem{coro}{\sc Corollary}

\newtheorem{nota}{\sc Notation}
\newtheorem{defin}{\sc Definition}
\newtheorem{rem}{\sc Remark}
\newtheorem{cla}{\sc Claim}
\newtheorem{ex}{\sc Example}

\newenvironment{remark}{\begin{rem}}{\hspace*{\fill}$\Diamond$\end{rem}}
\newenvironment{example}{\begin{ex}}{\hspace*{\fill}$\diamondsuit$\end{ex}}
\newenvironment{claim}{\begin{cla}}{\end{cla}}
\newenvironment{corollary}{\begin{coro}}{\end{coro}}

\newenvironment{definition}{\begin{defin}}{\end{defin}}

\title{Rate Distortion and Denoising of Individual Data 
Using Kolmogorov complexity}
\author{
Nikolai K. Vereshchagin\thanks{
NKV, Dept. Math. Logic \& Theor. Algor.,
Moscow State Univ., Russia. Email: nikolay.vereshchagin@gmail.com}
and
Paul M.B. Vit\'anyi\thanks{
PMBV, CWI, Science Park 123, 1098XG Amsterdam, the Netherlands.
Email: Paul.Vitanyi@cwi.nl}
}

\begin{document}
\maketitle
\begin{abstract}
We examine the structure of families of distortion balls from
the perspective of Kolmogorov complexity. Special attention is paid to
the canonical rate-distortion function of a source word
which returns the minimal Kolmogorov complexity of all distortion balls
containing that word subject to a bound on their cardinality. This canonical
rate-distortion function is related to the more standard
algorithmic rate-distortion function for the given distortion measure. 
Examples are given of list distortion, 
Hamming distortion, and Euclidean distortion.
The algorithmic rate-distortion function can behave
differently from Shannon's rate-distortion function.
To this end, we show that the canonical
rate-distortion function can and does assume a wide class of shapes
(unlike Shannon's); we relate low algorithmic mutual information
to low Kolmogorov complexity (and consequently suggest that certain aspects of the
mutual information formulation of Shannon's rate-distortion function
behave differently than would an analogous formulation using algorithmic
mutual information); we explore the notion that low Kolmogorov complexity
distortion balls containing a given word 
capture the interesting properties of that word
(which is hard to formalize in Shannon's theory) and this
suggests an approach to denoising; and, finally, we show that
the different behavior of the rate-distortion curves
of individual source words to some extent 
disappears after averaging over the source words.
\end{abstract}

\section{Introduction}
\label{sect.rdsf}
Rate distortion theory analyzes the transmission and 
storage of information at insufficient bit rates.
The aim is to minimize the resulting information loss
expressed in a given distortion measure.
The original data is called the `source word'
and the encoding used for transmission or storage
is called the `destination word.' The number of bits available
for a destination word is called the `rate.'
The choice of distortion
measure 
is usually a selection of which aspects of the source word are relevant
in the setting at hand, and
which aspects are irrelevant (such as noise). 
For example, in application to 
lossy compression of a sound file this results
in a compressed file where, among others, the very high and
very low inaudible frequencies have been suppressed. 
The distortion measure is chosen such that it penalizes 
the deletion of the inaudible
frequencies but lightly because they are not 
relevant for the auditory
experience. We study rate distortion of 
individual source words using Kolmogorov complexity and show
how it is related to
denoising.
The classical probabilistic theory is 
reviewed in Appendix~\ref{sect.ratedistortion}.
Computability notions are reviewed in Appendix~\ref{sect.computability}
and Kolmogorov complexity in Appendix~\ref{sect.kolmcompl}.
Randomness deficiency according to Definition~\ref{def.rd}
and its relation to the fitness of a destination word for
a source word is explained further in Appendix~\ref{sect.rd}. 
Appendix~\ref{sect.exhamming} gives the proof, required
for a Hamming distortion example, that
every large Hamming ball can be covered by a
small number of smaller
Hamming balls (each of equal cardinality).
More specifically, the number of covering balls 
is close to the ratio between the cardinality
of the large Hamming ball and the small Hamming ball.
The proofs of the theorems are deferred to Appendix~\ref{sect.proofs}.

\subsection{Related Work}
In \cite{Ko74} A.N. Kolmogorov formulated the
`structure function' which can be viewed as a proposal
for non-probabilistic model
selection. This function and the associated Kolmogorov
sufficient statistics are partially treated in
\cite{Sh83,Vy87,GTV01} 
and analyzed in detail in \cite{VV02}.
We will show that the structure function
approach can be generalized to give an approach to 
rate distortion and denoising of
individual data. 

Classical rate-distortion theory 
was initiated by Shannon in~\cite{Sh48}.
In~\cite{Sh59} Shannon gave a nonconstructive
asymptotic characterization of the expected rate-distortion curve of a
random variable
(Theorem~\ref{theo.shannon} in Appendix~\ref{sect.ratedistortion}).
References \cite{Be71,BG98} treat
more general distortion measures and random variables in the Shannon
framework.

References~\cite{YS93,MK94,SE03} relate
the classical and algorithmic approaches according to traditional
information-theoretic concerns. We follow their definitions of
the rate-distortion function.
The results show that if the source word is obtained from random
i.i.d. sources, then with high probability and in expectation
its individual rate-distortion curve is close to
the Shannon's single rate-distortion curve.
In contrast, our Theorem~\ref{theo.allshapesrd} shows that 
for distortion measures satisfying properties 1 through 4
below
there are many different shapes of individual 
rate-distortion functions related to the different
individual source words,
and many of them
are very different from Shannon's rate-distortion curve.

Also Ziv~\cite{Zi80} considers
a rate-distortion function for individual data.
The rate-distortion function is assigned to
every infinite sequence $\omega$ of letters of a finite alphabet $\Gamma$.
The source words $x$
are prefixes of $\omega$
and the encoding function is
computed by a finite state transducer.
Kolmogorov complexity is not involved.

In \cite{Sa94,Na95,CYV97,Do02} 
alternative approaches to denoising via compression 
and in \cite{RV06,rum} applications of the current work
are given.

 In \cite{VV02} Theorems~\ref{theo.allshapesrd}, \ref{th45} were obtained
 for a particular distortion measure relevant to model selection (the example 
${\cal L}$ in this paper).
The techniques used in that paper
do not generalize to prove the current theorems which concern 
arbitrary distortion measures
satisfying certain properties
given below.

\subsection{Results}
A source word is taken to be a finite binary string.
Destination words are finite objects (not necessarily finite binary strings).
For every destination word encoding a particular source word with
a certain distortion, there is a finite set of source words that are
encoded by this destination word with at most that distortion.
We call these finite sets of source words `distortion balls.'
Our approach is based on the Kolmogorov complexity           
of distortion balls. For every source word we
define its `canonical' rate-distortion function,
from which 
the algorithmic rate-distortion function of that source word 
can be obtained by a simple
transformation,
Lemma~\ref{lem.rg}. 

Below we assume that a distortion measure 
satisfies certain properties which are specified in the theorems
concerned.
In Theorem~\ref{theo.allshapesrd} it is shown that
there are distinct canonical rate-distortion curves (and hence distinct
rate-distortion curves) associated with 
distinct source words (although some curves may coincide). Moreover,
every candidate curve from a given family of curves is 
realized approximately as the 
canonical rate-distortion curve (and hence for a related family
of curves every  curve is realized approximately as the 
rate-distortion curve) of some
source word.
In Theorem~\ref{th-shannon-analog} we prove a Kolmogorov
complexity analogue for 
Shannon's theorem, Theorem~\ref{theo.shannon} 
in Appendix~\ref{sect.ratedistortion}, on the characterization
of the expected rate-distortion
curve of a random variable.
The new theorem states approximately the following:
For every source word and every destination word there exists
another destination word that has Kolmogorov complexity
equal to algorithmic information in the first destination word about the
source word, up to a logarithmic additive term,
and both destination words incur the same distortion
with the source word. (The theorem is given in the distortion-ball formulation
of destination words.)
In Theorem~\ref{th45} we show that, at every rate, 
the destination word incurring the least distortion
is in fact the `best-fitting' among all destination words at that rate. 
`Best-fitting' is taken in the sense of sharing the most
properties with the source word.
(This notion of a `best-fitting' destination word for a
source word can be expressed in Kolmogorov complexity, but 
not in the classic probabilistic framework. Hence there is no
classical analogue for this theorem.)
It turns out that this yields a method of denoising by compression.
Finally, in Theorem~\ref{thm.dresf}, we show that the expectation
of the algorithmic rate-distortion functions is
pointwise related to Shannon's rate-distortion function, where the closeness
depends on the Kolmogorov complexities involved and 
ergodicity and stationarity of the source.

\section{Preliminaries}

\subsection{Data and Binary Strings}
We write {\em string} to mean a finite binary string.
  Other finite objects can be encoded into strings in natural
ways.  The set of strings is denoted by $\{0,1\}^*$. The {\em length}
of a string $x$ is the number of bits in it denoted as $|x|$. The {\em empty}
string $\epsilon$ has length $|\epsilon| = 0$.
Identify the natural numbers 
${\cal N}$ (including 0) and $\{0,1\}^*$ according to the
correspondence 
 \begin{equation}\label{order}
 (0, \epsilon ), (1,0), (2,1), (3,00), (4,01), \ldots . 
 \end{equation}
Then, $|010|=3$.
The emphasis is on binary sequences only for convenience;
observations in every finite alphabet can be so encoded in a way
that is `theory neutral'. For example, if a finite alphabet $\Sigma$ has
cardinality $2^k$, then every element $i \in \Sigma$ can be encoded
by $\sigma(i)$ which
is a block of bits of length $k$. With this encoding every $x \in \Sigma^*$
satisfies that the Kolmogorov complexity 
$\K(x)=\K(\sigma(x))$ (see Appendix~\ref{sect.kolmcompl} for basic definitions
and results on Kolmogorov complexity) 
up to an additive constant that is
independent of $x$.

\subsection{Rate-Distortion Vocabulary}
Let ${\cal X}$ be a set, called  
the {\em source alphabet} whose elements are called 
{\em source words} or {\em messages}. 
We also use a set $\Y$ called the {\em destination alphabet},
whose elements are called {\em destination words}. 
(The destination alphabet is also called the reproduction alphabet.)
In general there are no restrictions on the set 
${\cal X}$; it can be countable or uncountable.
However, for technical reasons, we assume ${\cal X}= \{0,1\}^*$.
On the other hand, it is important that the set $\Y$ consists
of {\em finite objects}: we need that the notion of Kolmogorov complexity
$\K(y)$ be defined for all $y\in\Y$. 
(Again, for basic definitions and results on Kolmogorov complexity 
see Appendix~\ref{sect.kolmcompl}.) 
In this paper it is not essential
whether we use plain Kolmogorov complexity or the  prefix 
variant; we use plain Kolmogorov complexity.

Suppose we want to communicate a source word 
$x \in {\cal X}$ using a {\em destination word}
$y \in {\Y}$ 
that can be encoded in at most $r$ bits in the sense that
the Kolmogorov complexity $\K(y) \leq r$. 
Assume furthermore that we are given
a {\em distortion}
function
$d: {\cal X} \times {\Y} \rightarrow {\cal R} \bigcup \{\infty\}$,
that measures the fidelity of the destination word
against the source word.
Here ${\cal R}$ denotes the nonnegative real numbers,

\begin{definition}\label{def.rddr}
\rm
Let $x\in {\cal X} = \{0,1\}^*$ and ${\cal Q}$ denote the rational numbers.
The {\em rate-distortion function} $r_x: {\cal Q} \rightarrow {\cal N}$ is
the minimum number of bits in
a destination word $y$ 
to obtain a distortion of at most $\delta$ defined by
\[
r_x(\delta) = \min_{y \in {\Y}} \{\K(y) :  d(x,y)\le \delta\}
\]
The `inverse' of the above function is
is the {\em distortion-rate function} $d_x: {\cal N} \rightarrow {\cal R}$
 and is 
defined by
\[
d_x (r) = \min_{y \in {\Y}}  \{d(x,y) :    \K(y) \leq r \}.
\]
\end{definition}
These functions are analogs for individual source words $x$ of the
Shannon's rate-distortion
function defined in \eqref{eq.rndelta} and its related 
distortion-rate function, expressing
the least expected rate or distortion at which outcomes 
from a random source $X$ can be transmitted,
see Appendix~\ref{sect.ratedistortion}.

\subsection{Canonical Rate-Distortion Function}

Let ${\cal X}=\{0,1\}^*$ be the source 
alphabet, 
${\Y}$ a destination 
alphabet,
and $d$ a distortion measure.

\begin{definition}\label{def.distball}
\rm
A {\em distortion ball} $B(y,\delta)$ centered on $y \in {\Y}$
with radius $\delta\in\cal Q$ is defined by
\[
B(y,\delta)= \{x \in {\cal X}: d(x,y) \leq \delta \},
\]
and its cardinality is denoted by $b(y,\delta) = |B(y,\delta)|$.
(We will consider only pairs $(\Y,d)$
such that all distortion balls are finite.)
If the cardinality $b(y,\delta)$ depends only on
$\delta$ but not on the center $y$, then we denote it by  $b(\delta)$.
The family ${\A}^{d,\Y}$ is
defined as the set of all nonempty distortion balls.
The restriction 
to strings of length $n$ is denoted by 
${\A}^{d,\Y}_n$.
\end{definition}

To define the canonical rate-distortion function we need
the notion of the Kolmogorov complexity
of a finite set.

\begin{definition}\label{def.kcset}
\rm
Fix a computable 
total order on the set of all strings 
(say the order defined in \eqref{order}). 
The {\em Kolmogorov complexity $\K(A)$ of a finite set} 
is defined as the length of the shortest 
string $p$
such that the universal reference Turing machine $U$ 
given $p$ as input prints the list of all elements of $A$    
in the fixed order 
and halts. 
We require that the constituent elements are 
distinguishable so that we can tell
them apart. 
Similarly we define the {\em conditional} versions
$\K(A\mid z)$ and $\K(z\mid A)$ 
where $A$ is a finite set of strings
and $z$ is a string or a finite set of strings. 
\end{definition}

\begin{remark}
\rm
In Definition~\ref{def.kcset}
it is important that $U(p)$ halts after printing the last 
element in the list---in this way we know that the list is complete.
If we allowed $U(p)$ to not halt, then we would obtain the 
complexity of the so-called \emph{implicit description} of $A$, which can be
much smaller than $\K(A)$.
\end{remark}
\begin{remark}
\rm
We can allow  $U(p)$ to output the list of elements 
in any order in Definition~\ref{def.kcset}. This flexibility 
decreases $\K(A)$ 
by at most a constant not depending on $A$ but only depending
on the order in \eqref{order}.
The same applies to $\K(A\mid z)$.
On the other hand, if $A$ occurs in a conditional,
such as in $\K(z\mid A)$, then 
it {\em is} important that elements of $A$ are given in the fixed 
order. This is the case since the order in which the 
elements of $A$ are listed  
can provide extra information.
\end{remark}

\begin{definition}\label{def.Kfamily}
\rm
Fix a computable bijection $\phi$ from the family of all finite
subsets of $\{0,1\}^*$ to  $\{0,1\}^*$.
Let $\A$ be a finite family of finite subsets of ${\cal X}=\{0,1\}^*$.
Define the {\em Kolmogorov complexity} $\K(\A)$ 
by $\K(\A)=\K(\{\phi(A)): A\in\A\})$.
\end{definition}
\begin{remark}
\rm
An equivalent definition 
of $\K(A \mid z)$ and $\K(z \mid A)$ as in Definition~\ref{def.kcset}
is as follows. Let $\phi$ be as in Definition~\ref{def.Kfamily}.
Then we can define $\K(A \mid z)$ by $\K(\phi(A) \mid z)$ and  $\K(z \mid A)$ 
by $\K(z \mid \phi(A))$. 
\end{remark}

\begin{definition}\label{def.gx}
\rm
For every 
string $x$ 
the {\em canonical rate-distortion function} 
$g_x:\mathcal N\to\mathcal N$
is defined by
\[
g_{x}(l) = \min_{B \in {\A}^{d,\Y}} 
\{ \K(B) : x\in B,\log |B| \leq l\}.
\]
\end{definition}

In a similar way we can define
the \emph{canonical distortion-rate function}:
$$h_x(j)= \min_{B \in {\A}^{d,\Y}}\{\log|B|:
x\in B,\ C(B)\le j\}.
$$  

\begin{definition}
\rm
A {\em distortion family} ${\A}$ is a set of finite nonempty 
subsets of the set of source words
${\cal X}=\{0,1\}^*$. The restriction to source words of length $n$
is denoted by ${\A}_n$. 
\end{definition}

Every destination alphabet $\Y$ and 
distortion measure $d$ 
gives rise to a set of distortion balls 
${\A}^{d,\Y}$, which is a distortion family. Thus
the class of distortion families 
obviously includes every family of distortion
balls (or distortion spheres,
which is sometimes more convenient)
arising from every combination of 
destination set
and distortion measure.
It is easy to see that we also can 
substitute the more general distortion families ${\A}$
for ${\A}^{d,\Y}$ in the definitions
of the canonical rate-distortion and distortion-rate 
function.
In general,
the canonical rate-distortion function of $x$ can be quite different
from the rate-distortion function of $x$. However, by 
Lemma~\ref{lem.rg} below it turns out that  
for every distortion measure satisfying certain conditions
and for every $x$  
the rate-distortion function 
$r_x$ is obtained from $g_x$ by a simple transformation requiring
the cardinality of the distortion balls.

\begin{remark}
Fix a string $x\in\mathcal X=\{0,1\}^*$ 
and consider different distortion families $\A$.
Let $g_x^\A$ denote the canonical rate-distortion
function of $x$ with respect to a family $\A$.
Obviously, if $\A\subset\BB$ 
then $g_x^\A$ is pointwise not less than 
$g_x^\BB$ (and it may happen that $g_x^\A(i)\gg g_x^\BB(i)$ for some $i$). 
But as long as $\A$ satisfies certain natural properties, then
the set of all possible $g_x$, when
$x$ ranges over $\mathcal X$, does not depend on the particular $\A$ 
involved, see
Theorem~\ref{theo.allshapesrd}.
\end{remark}

\subsection{Use of the Big O Term}
In the sequel we use `additive constant $c$' or
equivalently `additive $O(1)$ term' to mean a constant.
accounting for the length of a fixed binary program,
independent from every variable or parameter in the expression
in which it occurs. 
Similarly we use 
`$O(f(m,n,\dots))$' to mean a function $g(m,n,\dots)$
such that $g(m,n,\dots) \leq c f(m,n,\dots)+c$ where $c$ 
is a fixed constant 
independent from every variable $m,n,\dots$ in the expression.

\section{Distortion Measures}


Since every family of distortion 
balls is a distortion family,
considering arbitrary distortion measures and destination alphabets 
results in distortion families. We consider
the following mild conditions on 
distortion families~${\A}$:
\begin{description}
\item{\bf Property 1.}
For every natural number $n$, 
the family ${\A}$ contains
the set $\{0,1\}^n$ of all strings of length $n$ as an element.
\item{\bf Property 2.}
All $x,y\in A\in {\A}$ satisfy 
$|x|=|y|$.
\item{\bf Property 3.}
Recall that ${\A}_n = \{A \in {\A}: A \subseteq  \{0,1\}^n \}$.
Then, $\K({\A}_n)=O(\log n)$.
\item{\bf Property 4.}
For every natural $n$, let
$\alpha_n$ denote the minimal number
that satisfies the following.
For every positive integer $c$ every 
set $A\in {\A}_n$ can be covered by at most
$\alpha_n |A|/c$ sets $B\in {\A}$ with $|B| \leq c$.
Call $\alpha_n$
the {\em covering coefficient} related to ${\A}_n$.
Property 4 is satisfied if $\alpha_n$ be bounded by
a polynomial in $n$.
The smaller the covering coefficient is, the more accurate will
be the description
that we obtain of the shapes of the structure functions below.
\end{description}
The following three example
families ${\A}$ satisfy all four properties.
\begin{example} \label{exam.list}
\rm
${\cal L}$ {\em the list distortion family}. 
Let ${\cal L}_n$ 
be the family of all nonempty subsets
of $\{0,1\}^n$. 
This is the family of distortion balls 
for list distortion, which we define as follows.
Let 
${\cal X} =\{0,1\}^*$ and 
${\Y}=\bigcup_n\mathcal L_n$.
A  source word $x \in \{0,1\}^n$ is 
encoded by a destination word
which is a subset or {\em list} 
$S \subseteq \{0,1\}^n$ with $x \in S$.
Given $S$, we can retrieve $x$ by its index of $\log |S|$ bits in $S$,
ignoring rounding up, whence the name `list code.'
The distortion measure is $d(x,S)= \log |S|$ if $x \in S$,
and $\infty$ otherwise. Thus, distortion balls come only in the form 
$B(S,\log |S|)$ with cardinality $b(S,\log |S|)=|S|$.
Trivially, the covering coefficient 
as defined in property~4,
for the list distortion family ${\cal L}$,
satisfies $\alpha_n \leq 2$.
Reference~\cite{VV02} describes
all possible canonical distortion-rate curves, called  
Kolmogorov's  structure function there and first defined in \cite{Ko74}.
The distortion-rate function for list distortion 
coincides with the canonical distortion-rate function.
The rate-distortion
function of $x$ for list distortion is
\[
r_x(\delta) = 
\min_{S \subseteq \{0,1\}^n} \{\K(S): x \in S , \; \log |S| \leq \delta \}
\]
and essentially coincides with the canonical rate-distortion function
($g_x$ is the restriction of $r_x$ to $\cal N$). 
\end{example}

\begin{example}
\rm
${\cal H}$ {\em the Hamming distortion family}. 
Let ${\cal X} = {\Y} =\{0,1\}^*$.
A source word  $x \in \{0,1\}^n$ is
encoded by a destination word $y \in \{0,1\}^n$. 
For every positive integer $n$, the {\em Hamming distance}
between two strings $x= x_1 \ldots x_n$ and
$y =y_1 \ldots y_n$ is defined by 
\begin{equation}\label{eq.hamdist}
d(x,y)= \frac{1}{n} |\{i : x_i\neq y_i\}|.
\end{equation}
If $x$ and $y$ have different lengths, then $d(x,y)=\infty$.
A {\em Hamming ball} in $\{0,1\}^n$ with center
$y\in \{0,1\}^n$ and radius $\delta$ ($0 \leq \delta \leq 1$)  is the set 
$B(y,\delta)=\{x\in\{0,1\}^n: d(x,y)\le \delta \}$.
Every $x$ is in either $B( 00\ldots 0,\frac{1}{2})$ or
$B(11\ldots 1,\frac{1}{2})$, so we need to consider only 
Hamming distance $0 \leq \delta \leq \frac{1}{2}$.
Let ${\cal H}_n$ be the family of all Hamming balls
in $\{0,1\}^n$. 
We will use the following
approximation of $b(\delta)$---the cardinality of Hamming balls 
in ${\cal H}_n$ of radius 
$\delta$.
Suppose that $0 \le \delta \le \frac{1}{2}$ and $\delta n$ is an integer,
and let
$H(\delta)=\delta\log 1/\delta+(1-\delta)\log1/(1-\delta)$
be Shannon's binary entropy function. Then,
\begin{equation}
\label{binom-entropy}
2^{n H(\delta)-\log n/2-O(1)} \leq
b(\delta) \leq 2^{nH(\delta)}.
\end{equation}
In Appendix~\ref{sect.exhamming} 
it is shown that the covering coefficient 
as defined in property~$4$,
for the Hamming distortion family ${\cal H}_n$,
satisfies $\alpha_n = n^{O(1)}$. The function
\[
r_x(\delta) = \min_{y \in \{0,1\}^n} \{\K(y): 
 d(x,y) \leq \delta  \}
\]
is the rate-distortion
function of $x$ for Hamming distortion. An approximation to
one such function is depicted in Figure~\ref{ham.eps}.
\end{example}

\begin{example}
\rm
${\cal E}$ {\em the Euclidean distortion family}. 
Let ${\cal E}_n$ be 
the family of all intervals in $\{0,1\}^n$,
where an interval  is a  
subset of $\{0,1\}^n$ of the form $\{x: a\leq x\leq b\}$
and $\leq$ denotes the lexicographic ordering on $\{0,1\}^n$.
Let ${\Y} =\{0,1\}^*$.
A  source word $x \in \{0,1\}^n$ is
encoded by a destination word $y \in \{0,1\}^n$. 
Interpret strings in $\{0,1\}^n$ as
binary notations for rational numbers in the segment $[0,1]$.
Consider the Euclidean distance $|x-y|$
between rational numbers $x$ and $y$.
The balls in this metric are intervals;
the cardinality of a ball of radius $\delta$
is about $\delta 2^n$.
Trivially, the covering coefficient 
as defined in property~$4$,
for the Euclidean distortion family ${\cal E}_n$,
satisfies  $\alpha_n \leq 2$.
The function
\[
r_x(\delta) = \min_{y \in \{0,1\}^n} \{ \K(y):  |x-y| \leq \delta \}
\]
is the rate-distortion
function of $x$ for Euclidean distortion.
\end{example} 
All the properties 1 through 4 
are straightforward for all three families, 
except property~$4$ in the case
of the family of Hamming balls.

\section{Shapes}\label{sec1}

The rate-distortion functions of the 
individual strings of length $n$ can assume roughly
every shape. That is, every shape 
derivable from a function in the large family
$G_n$ of Definition~\ref{def.gx} below through transformation
\eqref{eq.sfrd}.

We start the formal part of this section. 
Let ${\A}$ be a distortion family satisfying
properties~1 through~4.

Property $1$ implies that $\{0,1\}^n \in {\A}$ and property $4$ 
applied to $\{0,1\}^n$ and $c=1$,
for every $n$, implies trivially that
the family ${\A}$ contains the singleton set
$\{x\}$ for every $x\in\{0,1\}^*$. Hence,
$$
g_x(0)= \K(\{x\})= \K(x)+O(1).
$$ 
Property~$1$
implies that for every $n$ and string $x$ of length $n$,
\[
g_x(n)\leq \K(\{0,1\}^n)=\K(n)+O(1)\leq \log n+O(1).
\]
Together this means that for every $n$ and every
string $x$ of length $n$,  
the function $g_x(l)$ decreases from about $\K(x)$
to about $0$ as $l$ increases from 0 to $n$.

\begin{lemma}\label{lem.shapesg}
Let ${\A}$ be a distortion family satisfying
properties~$1$ through $4$.
For every $n$ and every string $x$ of length $n$ we have
$g_x(n)= O(\log n)$, and
$0\le g_x(l)-g_x(m)\leq m-l+O(\log n)$
for all $l<m\leq n$. 
\end{lemma}
\begin{proof}
The first equation and the left-hand inequality of
the second equation are
straightforward.
To prove
the right-hand inequality
%
let $A$ witness $g_x(m)=k$, which implies that 
$\K(A)=k$ and $ \log |A|\leq m$. By Property 4 there is 
a covering of $A$ by at most $\alpha_n |A|/2^{l}$ sets in ${\A}_n$
of cardinality at most $2^{l}$ each. 
Given a list of $A$ and a list of $\A_n$, we can find 
such a covering.
Let $B$ be one of 
the covering sets containing $x$.
Then, $x$ can be specified by $A,n,l,\A_n$ 
and the index $i$ of $B$
among the covering sets.
We need also $O(\log k+\log\log i+\log\log l +\log \log n)$
extra bits to separate the descriptions of $A$ and $\A_n$, and 
the binary representations of $i,n,l$, from one another.
Without loss of generality we can assume that $k$
is less than $n$.
Thus all the extra information
and separator bits are included in $O(\log n)$ bits.
Altogether,
$\K(B)\leq \K(A) +m-l +O(\log n)\leq k +m-l +O(\log n)$, which shows
that $g_x(l)\le k+m-l+O(\log n)=g_x(m)+m-l+O(\log n)$.
\end{proof}

\begin{example}\rm
Lemma~\ref{lem.shapesg} shows
that  
$$
\K(x)-i-O(\log n)\leq g_x(i)\leq n-i+O(\log n),
$$
for every $0 \leq i \leq n$.
The right-hand inequality 
is obtained by setting $m=n$, $l=i$ in
the lemma, yielding
$$
g_x(i)=g_x(i)-g_x(n)+O(\log n)\leq n-i+O(\log n).
$$
The left-hand inequality 
is obtained by setting $l=0$, $m=i$ 
in the lemma, yielding
$$
\K(x)-g_x(i)=g_x(0)-g_x(i) +O(1)\le i-0+O(\log n).
$$
The last displayed equation can also be shown by a simple direct argument:
$x$ can be described by the minimal description
of the set $A \in {\A}$ 
witnessing $g_x(i)$ and by the ordinal number of $x$ in $A$.
\end{example}

The rate-distortion
function $r_x$ differs
from $g_x$ by just a change of scale depending on the distortion family
involved, provided certain computational requirements are fulfilled.
See Appendix~\ref{sect.computability} for computability notions. 

\begin{lemma}\label{lem.rg}
Let  ${\cal X} = \{0,1\}^*$, ${\Y}$, and $d$, be the 
source alphabet, destination alphabet,
and distortion measure, respectively. 
Assume that the set 
$\{\pair{x,y,\delta}\in\mathcal X\times\Y\times\Q: d(x,y)\le \delta\}$
is decidable; that $\Y$ is recursively enumerable; and  
that for every $n$ the cardinality 
of every ball in ${\A}^{d,\Y}_n$ of radius $\delta$ is at most 
$b_n(\delta)$ and at least $b_n(\delta)/\beta(n)$, where 
$\beta(n)$ is polynomial in $n$ and $b_n(\delta)$ is a function
of $n,\delta$; and that the distortion family $\A^{d,\Y}$
satisfies properties 1 through 4.  
Then, for every $x\in\{0,1\}^n $ and every rational $\delta$
we have 
\begin{equation}\label{eq.sfrd}
r_x (\delta ) = g_x(\lceil \log b_n(\delta) \rceil)+O(\K(\delta)+\log n).
\end{equation}
\end{lemma}
\begin{proof}
Fix $n$ and a string $x$ of length $n$.
Consider the auxiliary function
\begin{equation}\label{eq.tilde}
\tilde r_x(\delta) = \min_{y\in \Y} \{\K(B(y,\delta)): 
 d(x,y) \leq \delta  \}.
\end{equation}
We claim that 
$\tilde r_x(\delta)= r_x(\delta)+O(\K(\delta)+\log n)$.
Indeed, let $y$ witness $r_x(\delta)=k$. 
Given $y,\delta,n$ we can compute
a list of elements of the ball $B(y,\delta)$: for all strings 
$x'$ of length $n$ determine whether $d(x',y)\le\delta$. 
Thus $\K(B(y,\delta))<k+O(\K(\delta)+\log n)$, hence 
$\tilde r_x(\delta)<k+O(\K(\delta)+\log n)$.
Conversely, let $B(y, \delta)$ witness $\tilde r_x(\delta)=k$. 
Given a list of the elements of $B(y,\delta)$ and $\delta$
we can recursively enumerate ${\Y}$ to find the first element
$y'$ with $B(y',\delta)=B(y,\delta)$ (for every enumerated $y'$ compute 
the list $B(y',\delta)$ and compare it to the given list $B(y,\delta)$). 
Then,
$\K(y')\le k+O(\K(\delta))$ and $d(x,y')\le\delta$.
Hence $r_x(\delta)<k+O(\K(\delta))$.

Thus, it suffices to show that 
\[
\tilde r_x (\delta ) = g_x(\lceil \log b_n(\delta) \rceil)+O(\log n).
\]

($g_x(\lceil \log b_n(\delta) \rceil)\leq\tilde r_x (\delta)$)
Assume $\tilde r_x(\delta)=k$ is witnessed by a distortion ball $B(y, \delta)$.
By our assumption, the  cardinality of $B(y,\delta)$ is at most 
$b_n(\delta)$, and hence $g_x(\lceil \log b_n(\delta) \rceil ) \leq k$. 

($\tilde r_x (\delta) \leq g_x(\lceil \log b_n(\delta) \rceil)+O(\log n)$)
By Lemma~\ref{lem.shapesg},
$g_x(l)$ and $g_x(l-m)$ differ by at most $m+O(\log n)$. 
Therefore it suffices to show that 
$\tilde r_x (\delta) \leq g_x(\lceil \log b_n(\delta) \rceil-m)$
for some $m=O(\log n)$. We claim that this happens for
$m=\lceil\log\beta(n)\rceil+1$. Indeed, let  
$g_x(\lceil \log b_n(\delta) \rceil-m)=k$ be witnessed
by a distortion ball $B$. Then, 
$|B|\le 2^{\lceil\log b_n(\delta)\rceil}/(2\beta(n))<
b_n(\delta)/\beta(n)$.
This implies that the radius of $B$ is less than $\delta$
and hence $B$ witnesses $\tilde r_x (\delta)\le k$. 
\end{proof}

\begin{remark}\label{rem.logn} 
\rm
When measuring distortion we usually do 
not need rational numbers with numerator or denominator more
than $n=|x|$. Then, the term $O(C(\delta))$ in \eqref{eq.sfrd}
is absorbed by the term $O(\log n)$. 
Thus, describing the family of $g_x$'s  we obtain an approximate
description of all possible rate-distortion functions $r_x$ for 
given destination alphabet and distortion measure, satisfying the computability 
conditions, by using the transformation \eqref{eq.sfrd}.
An example of an approximate
rate-distortion curve $r_x$ for some string $x$
of length $n$ for Hamming distortion is given in Figure~\ref{ham.eps}.
\end{remark}
\begin{remark} 
\rm
The computability properties of the functions
$r_x$, $d_x$, and $g_x$, as well as the relation between
the destination word for a source word and the related distortion ball, is
explained in Appendix~\ref{sect.computability}.
\end{remark}

We present an approximate 
description of the family of possible $g_x$'s below. It turns
out that the description does not depend on the particular distortion family
$\A$ as long as properties 1 through 4 are satisfied.

\begin{definition}
\rm
Let $G_n$ stand for the class of all
functions $g:\{0,1,\dots,n\}\rightarrow {\cal N}$ such
that $g(n)=0$  and
$g(l-1)\in\{g(l),g(l)+1\}$ for all
$1\leq l \leq n$.
\end{definition}

In other words, a function $g$ is in $G_n$ iff
it is nonincreasing and the function $g(i)+i$
is nondecreasing and $g(n)=0$.
The following result is a generalization to
arbitrary distortion measures of Theorem IV.4
in \cite{VV02}
dealing with $h_x$ (equaling $d_x$ in the particular case
of the distortion family 
${\cal L}$). There, the precision in Item (ii) for source words of length $n$
is $O(\log n)$, rather than the $O(\sqrt{n \log n})$ we obtain
for general distortion families.

\begin{theorem}\label{theo.allshapesrd}
Let ${\A}$ be a distortion family satisfying
properties~$1$ through~$4$.

{\rm (i)}  For every $n$ and every string $x$ of length $n$, the function
$g_x(l)$ is equal to $g(l)+O(\log n)$ for some function $g \in G_n$.

{\rm (ii)}
Conversely, for every $n$ and every function $g$ in $G_n$,
there is a string
$x$ of length $n$ such that for every $l=0,\dots,n$,
$g_x(l)=g(l)+O(\sqrt{n\log n})$.
\end{theorem}

\begin{remark}
\rm
For fixed $k \leq n$ the number of different integer functions $g \in G_n$ 
with
$g(0) = k$ 
is ${n \choose k}$. 
For $k=\frac{1}{2}n$,
this number is of order $2^n/\sqrt{ n}$,
and therefore far greater than the number
of strings $x$ of length 
$n$ and Kolmogorov complexity 
$\K(x) = k = \frac{1}{2}n$ which is at most $2^{n/2}$.                   
This explains the fact that in Theorem~\ref{theo.allshapesrd}, Item (ii),
we cannot precisely match a string $x$ of length $n$ to
every function $g \in G_n$, and therefore have to use approximate
shapes.
\end{remark}

\begin{example}
\rm
By Theorem~\ref{theo.allshapesrd}, Item (ii), for every $g \in G_n$ 
there is a string $x$ of length $n$ that has $g$ for its canonical
rate-distortion function $g_x$ up to an additive $O(\sqrt{n \log n})$ term. 
By \eqref{binom-entropy}, \eqref{eq.sfrd}, and Remark~\ref{rem.logn},
$$
r_x(\delta)=
g_x(nH(\delta))+O(\log n),
$$
for $0 \leq \delta \leq \frac{1}{2}$.
\begin{figure}[ht]
\begin{center}
\epsfxsize=3.5in
\leftline{\hskip8pc\epsfbox{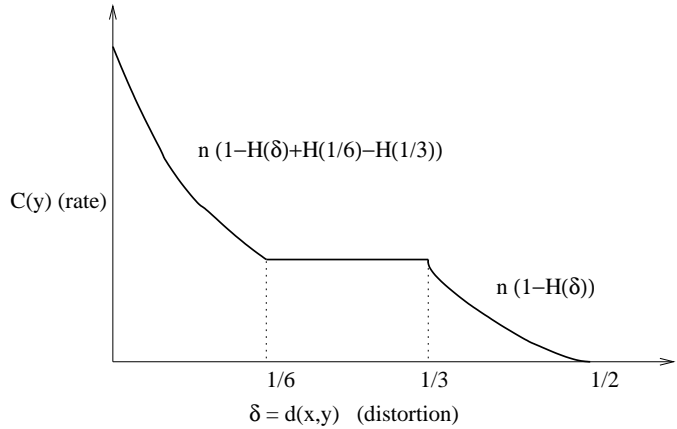}}
\end{center}
\caption{An approximate rate-distortion function for Hamming distortion}
\label{ham.eps}
\end{figure}
Figure~\ref{ham.eps} gives the graph of a particular function 
$r(\delta) = g(nH(\delta))$ with $g$ defined as follows:
 $g(l) = n(1+H(\frac{1}{6})-H(\frac{1}{3}))-l$
for $0 \leq l \leq nH(\frac{1}{6})$, 
$g(l)=n(1+H(\frac{1}{6})-H(\frac{1}{3}))$ for 
$nH(\frac{1}{6}) < l \leq nH(\frac{1}{3})$,
and $g(l)=n-l$ for $nH(\frac{1}{3}) < l \leq n$. 
In this way, $g \in G_n$.
Thus, there is a string $x$ of length $n$ with its rate-distortion
graph $r_x (\delta)$ 
in a strip of size $O(\sqrt{n\log n})$ around the
graph of $r(\delta)$. Note that $r_x$ is almost constant on
the segment $[ \frac{1}{6}; \frac{1}{3}]$.
Allowing the
distortion to increase on this interval, all the way from
$\frac{1}{6}$ to $ \frac{1}{3}$, so allowing $n/6$ incorrect extra
bits, we still cannot significantly decrease the rate.
This means that the distortion-rate function $d_x(r)$
of $x$ drops from $\frac{1}{3}$ to $\frac{1}{6}$ 
near the point $r=n(1-H(\frac{1}{3}))$,
exhibiting a very unsmooth behavior.
\end{example}

\section{Characterization}

Theorem~\ref{th-shannon-analog} below states that a destination word that
codes a given source word and
 minimizes the algorithmic mutual information with 
the given source word gives no
advantage in rate 
 over a minimal Kolmogorov complexity destination word that codes the source word.
This 
theorem
can be compared with Shannon's theorem, Theorem~\ref{theo.shannon} in
Appendix~\ref{sect.ratedistortion}, about
the expected rate-distortion curve of a random variable.

\begin{theorem}    \label{th-shannon-analog}
Let ${\A}$ be a distortion family 
satisfying properties~$2$
and~$3$, 
and
 ${\A}(x) = \{A  \in {\A}: x \in A\}$.
For every $n$ and string $x$ of length $n$ and every $B \in {\A}(x)$
there is an $A \in {\A}(x)$ with 
$\lceil \log |A|\rceil =\lceil \log |B| \rceil$ and
$\K(A)\leq I(x:B)+O(\log \K(B)+\log n)$,
where $I(x:B)=\K(B)-\K(B\mid x)$ stands for the
algorithmic information in $x$ about $B$.
\end{theorem}

For further information about $I(x:B)$ see Definition~\ref{def.mi} in
Appendix~\ref{sect.kolmcompl}.
The proof of Shannon's theorem, Theorem~\ref{theo.shannon},
and the proof of the current theorem are very different.
The latter proof uses techniques 
that may be
of independent interest.
In particular, we use an online 
set cover algorithm where the sets come sequentially and we always have
to have the elements covered that occur in a certain number of sets,
Lemma~\ref{th5} in Appendix~\ref{sect.proofs}.

\begin{example}
\rm
Theorem~\ref{th-shannon-analog} states that
for an appropriate distortion family ${\A}$ of nonempty finite subsets
of $\{0,1\}^*$ 
and for every string $x \in \{0,1\}^*$, if there exists an $A\in {\A}$
of cardinality
$2^l$ or less
containing $x$ that has small algorithmic information about $x$,
then there exists another
set $B\in {\A}$ containing $x$ that has also at most $2^l$ elements
and has small
Kolmogorov complexity itself.
For example, in the case of Hamming distortion, if for a given string $x$
there exists a string $y$ at Hamming distance
$\delta$ from $x$
that has small information about $x$, then there exists another
string $z$ that is also within distance $\delta$ of  $x$ and has small
Kolmogorov complexity itself (not only small algorithmic 
information about $x$).
\end{example}

\section{Fitness of Destination Word}\label{sect.fitness}

In Theorem~\ref{th45} we show that if a destination word 
of a certain maximal Kolmogorov complexity
has minimal distortion with respect to the source word, then it
also is the (almost) best-fitting destination word in the sense 
(explained below)
that
among all destination words of that Kolmogorov complexity
it has the most properties in common with the
source word.
`Fitness' of individual strings to an individual
destination word is hard, if not impossible, to describe
in the probabilistic framework. However, for the combinatoric
and computational notion of Kolmogorov complexity it is natural to describe
this notion using `randomness deficiency' as in Definition~\ref{def.rd} below.

Reference \cite{VV02} uses `fitness' 
with respect to the particular distortion family
${\cal L}$. We briefly overview the generalization to arbitrary
distortion families satisfying properties 2 and 3 (details,
formal statements and proofs about ${\cal L}$ can be found in the 
cited reference).
The goodness of fit of a destination word $y$ for
a source word $x$ with respect to an arbitrary distortion family ${\A}$
is defined by the randomness deficiency of $x$ in the
the distortion ball $B(y, \delta)$ with $\delta=d(x,y)$. 
The lower the randomness deficiency, the better is the fit.
\begin{definition}\label{def.rd}
\rm
The {\em randomness deficiency} of $x$ in a set $A$ with $x \in A$
is defined as $\delta (x \mid A) = \log |A| - \K(x\mid A)$.
If $\delta (x \mid A)$ is small then $x$ is a {\em typical} element of $A$.
Here `small' is taken as $O(1)$ or $O(\log n)$ where $n=|x|$, 
depending on the context of the future statements.
\end{definition}

The randomness deficiency can be little smaller
than 0, but not more than
a constant.
\begin{definition}
\rm
Let $\beta$ be an integer parameter and $P \subseteq A$.
We say $P$ is a {\em property} in 
$A$ if $P$ is a `majority' subset of
$A$, that is,  $|P| \geq (1-2^{\beta})|A|$. We say that 
$x \in A$ \emph{satisfies} property $P$ if 
$x \in P$.
\end{definition}

If the randomness deficiency $\delta(x \mid A)$ is not much greater than 0,
then there are no simple special properties that
single $x$ out from the majority of strings to be drawn from $A$.
This is not just terminology: 
If $\delta (x  |  A)$ is small enough,
then $x$ satisfies {\em all} properties of low Kolmogorov complexity
in $A$ (Lemma~\ref{lemma.property} in Appendix~\ref{sect.rd}).
If $A$ is a set containing $x$ such that $\delta(x \mid A)$ is 
small 
then we say that $x$ is
a set of good fit for $x$.
In \cite{VV02} the notion of 
models for $x$ is considered: Every finite set of strings
containing $x$ is a {\em model} for $x$. 
Let $x$ be a string of length $n$ and choose an integer $i$ 
between 0 and $n$. Consider models for $x$ of 
Kolmogorov
complexity at most $i$.
Theorem~IV.8 and Remark IV.10 in \cite{VV02}
show
for the distortion family ${\cal L}$
that $x$ has minimal
randomness deficiency in every set that witnesses $h_x(i)$ 
(for ${\cal L}$ we have $h_x(i)=d_x(i)$),
ignoring additive $O(\log n)$ terms. That is, up to the stated precision
every such witness set is the best-fitting model that is
possible at model Kolmogorov complexity at most $i$. 
 It is
remarkable, and in fact unexpected to the authors,
that the analogous result
holds for arbitrary distortion families provided 
they satisfy properties 2 and 3.

\begin{theorem}\label{th45}
Let ${\A}$ be a distortion family 
satisfying properties~$2$ and~$3$ 
and $x$ a string of length $n$. 
Let $B$ be a set in $\A$ with 
$x \in B$.
Let $A_x$ be a set
of minimal Kolmogorov complexity
among the sets $A\in{\A}$ with $x\in A$ and 
$\lceil \log |A| \rceil= \lceil \log |B| \rceil$.
Then,
\[
\K(A_x)+\log |A_x|-\K(x)\leq
\delta(x \mid B)
+O(\log \K(B)+ \log n).
\]
\end{theorem}
\begin{lemma}\label{lemma.deltaab}
For every set $A$ with  $x \in A$,
\begin{equation}\label{eq.deltaab}
\K(A)+\log |A|-\K(x) \ge\delta (x \mid A),
\end{equation}
up to a  $O(\log n)$ additive term. 
\end{lemma}
\begin{proof}
The inequality \eqref{eq.deltaab}
means that that 
$$\K(A)+\log |A|-\K(x) \ge \log |A|-\K(x\mid A)+O(\log n),$$
that is,
$\K(x)\le \K(A)+\K(x\mid A)+O(\log n)$.
The latter inequality follows 
from the general inequality 
$\K(x)\le \K(x,y) \leq \K(y)+\K(x\mid y)+O(\log\K(x\mid y))$,  
where $\K(x\mid y)\le\K(x)+O(1)\le n+O(1)$.
\end{proof}

A set $A$ with $x \in A$ is an algorithmic {\em sufficient statistic} 
for $x$ if
$\K(A)+\log |A|$ is close to $\K(x)$.
Lemma~\ref{lemma.deltaab} shows that every sufficient statistic for $x$ is
a model of a good fit for $x$.

\begin{example}\label{th44}
\rm
Consider the elements of every $A\in {\A}$ uniformly distributed.
Assume that we are given a string $x$ that was 
obtained by a random sampling
from an unknown set $B\in {\A}$ 
satisfying $\K(B)\le n=|x|$.
Given $x$
we want to recover $B$, or some $A\in {\A}$ that
is ``a good hypothesis to be the source of $x$'' in the sense
that the randomness deficiency $\delta (x \mid A)$ is small. 
Consider the set $A_x$ from  Theorem~\ref{th45} as such
a hypothesis. We claim that 
with high probability $\delta(x \mid A_x)$ is of order $O(\log n)$.
More specifically, for every $\beta$ the probability of the event
$\delta(x \mid A_x)>\beta$
is less than 
$2^{-\beta+O(\log n)}$, 
which is negligible for $\beta=O(\log n)$. 
Indeed, 
if $x$ is chosen uniformly  at random in $B$, then
with high probability 
(Appendix~\ref{sect.rd})
the randomness deficiency $\delta (x \mid B)$ is small.
That is, with probability more than $1-2^{-\beta}$ 
we have $\delta(x \mid B)\le\beta$.
By Theorem~\ref{th45} and \eqref{eq.deltaab}
we also have $\delta(x \mid A_x)\le\delta(x \mid B)+O(\log n)$.
Therefore the probability of the event
$\delta(x \mid A_x)>\beta$
is less than 
$2^{-\beta+O(\log n)}$.
\end{example}

\begin{example}
\rm
Theorem~\ref{th45} says that for fixed
log-cardinality $l$ the model that has minimal Kolmogorov complexity has
also minimal randomness
deficiency among models of that log-cardinality.
Since $g_x$ satisfies  Lemma~\ref{lem.shapesg}, we have also that for every
$k$ the model of Kolmogorov complexity at most
$k$ that minimizes the log-cardinality also minimizes randomness
deficiency among models of that Kolmogorov complexity. 
These models can be computed in the limit, in the first case
by running all programs up to $k$ bits and always keeping the one
that outputs the smallest set in ${\A}$ containing $x$, and in the second case
by running all programs up to $n=|x|$ bits and always keeping the
shortest one that outputs a set in ${\A}$ containing $x$
having log-cardinality at most $l$.
\end{example}

\section{Denoising}

In Theorem~\ref{th45} using \eqref{eq.deltaab} we obtain
\begin{equation}\label{eq.dAB}
\delta(x \mid A_x)\le \delta(x \mid B)+O(\log \K(B)+\log n).
\end{equation}
This gives a method
to identify good-fitting models for $x$ using compression, as follows. 
Let $k= \K(A_x)$ and $l= \lceil \log |B| \rceil$.
If $A_x$ is a
set of minimal Kolmogorov complexity
among sets  $A \in {\A}$ with $x\in A$ and $ \lceil \log |A| \rceil=l$,
then by \eqref{eq.dAB}
the hypothesis ``$x$ is chosen at random
in $A_x$'' is (almost) at least as plausible as
the hypothesis ``$x$ is chosen at random
in $B$'' for every simply described
$B\in {\A}$ 
(say, $\log \K(B)=O(\log n)$) 
with  $ \lceil \log |B| \rceil=l$.

Let us look at an example
of denoising by compression
(in the ideal sense of Kolmogorov complexity) for Hamming distortion.
Fix a target string $y$ of length $n$ and a 
distortion $0 \leq \delta \leq \frac{1}{2}$.
(This string $y$ functions as the destination word.)
Let a string $x$ be a noisy version of
$y$ by changing at most $n\delta$ randomly chosen bits in $y$
(string $x$ functions as the source word).
That is,
the string $x$ is chosen uniformly at random in the Hamming ball
$B=B(y,\delta)$.
Let $\hat{x}$ be 
a string witnessing 
$r_x(\delta)$, that is, $\hat{x}$ is a string
of minimal Kolmogorov complexity  with $d(x,\hat{x}) \leq \delta$
and $r_x(\delta)=C(\hat{x})$.
We claim that at distortion $\delta$ the string
 $\hat{x}$ is a good candidate for
a denoised version of $x$, that is, the target string $y$.
This means that
in the two-part description
$(\hat{x},\hat{x} \oplus x)$
of $x$, the second part (the bitwise XOR of $x$ and $\hat{x}$)
is noise: 
$\hat{x} \oplus x$ is a random string 
in the Hamming ball $B(00\dots0,\delta)$ in the sense 
that $\delta(\hat{x} \oplus x \mid B(00\dots0,\delta))$ is negligible.
Moreover, even the conditional Kolmogorov complexity
$\K(\hat{x} \oplus x \mid \hat x)$ is close to $\log b(\delta)$.

Indeed, 
let $l=\lceil\log|B|\rceil$.
By Definition~\ref{def.gx} of $g_x$, 
Theorem~\ref{th45} implies that
$$
g_x(l)+l-\K(x)\le \delta(x \mid B),
$$
ignoring additive terms of $O(\log n)$
and observing that the additive 
term $\log \K(B)$ is absorbed by $O(\log n)$.
For every $x$,
the rate-distortion function $r_x$ of $x$ differs from
$g_x$ just by changing the scale of the argument as in \eqref{eq.sfrd}.
More specifically,
we have
$r_x(\delta) = g_x(l)$ and hence
\[
r_x(\delta)+l-\K(x)\leq \delta(x \mid B).
\]
Since we assume that $x$ is chosen uniformly
at random in $B$, the randomness deficiency 
$\delta(x \mid B)$ is small, say $O(\log n)$ with high probability.
Since 
$r_x(\delta)=\K(\hat{x})=\K(B(\hat{x},\delta))+O(\K(\delta))$,
$\K(\delta)=O(\log n)$, and $l=\lceil\log b(\delta)\rceil$,
it follows that with high probability, and the equalities up to an
additive $O(\log n)$ term,
$$
0 =  \K(\hat{x})+l- \K(x)= \K(B(\hat{x},\delta))+
\log b(\delta)-\K(x).
$$
Since by construction $x \in B(\hat{x},\delta)$, 
the displayed equation shows that 
the ball $B(\hat{x},\delta)$ is a sufficient statistic for $x$.
This implies that $x$ is a typical element of $B(\hat{x},\delta)$,
that is, $\K(x\oplus\hat x \mid \hat{x})=\K(x \mid \hat{x})=
\K(x \mid B(\hat{x},\delta),p)$ 
is close to $\log b(\delta)$.
Here $p$ is an appropriate
program of $O(\C(\delta))=O(\log n)$ bits.

This provides a method of denoising via compression, 
at least in theory.
In order to use the method practically, admittedly with a leap of faith,
we ignore the ubiquitous $O(\log n)$ additive terms,
and use real compressors to
approximate the Kolmogorov complexity, similar to what was done in  
\cite{Li01,Li04}.
The Kolmogorov complexity is not computable and can be approximated
by a computable process from above but not from below, while a real 
compressor is computable. Therefore, the approximation of the Kolmogorov 
complexity by a real compressor involves for some arguments errors that can 
be high and are in principle unknowable. Despite all these caveats it turns
out that the practical analogue of the theoretical method works surprisingly
well in all experiments we tried \cite{RV06}. 

\begin{figure}
\begin{center}
\epsfxsize=3.5in
\leftline{\hskip8pc\epsfbox{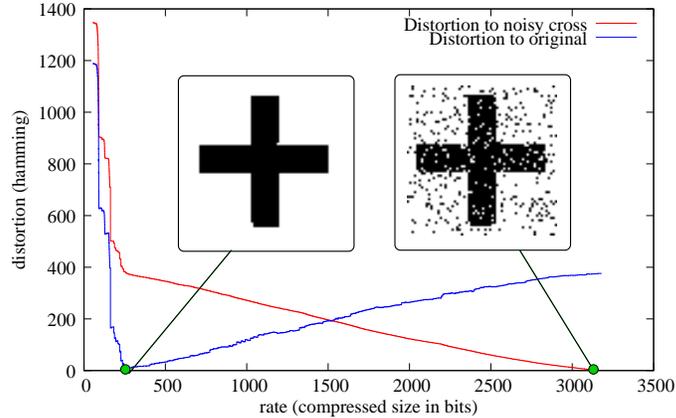}}
\end{center}
\caption{Denoising of the noisy cross}
\label{fig:cross}
\end{figure}

As an example, we approximated the distortion-rate
function of a noiseless cross of very low
Kolmogorov complexity, to which artificial noise was added to obtain
a noisy cross,  \cite{RV06}. 
Figure~\ref{fig:cross} shows two graphs. The first graph, hitting
the horizontal axis at about 3100 bits, denotes the Hamming distortion
on the vertical axis of the best
model for
the noisy cross with respect to the original noisy cross
 at the rate given on the horizontal axis.
The line hits zero distortion at model cost
bit rate about 3100,
when the original noisy cross is retrieved. The best model of the noisy cross
at this rate, actually the original noisy cross, 
is attached to this point. The second graph, hitting the horizontal axis at
about 250 bits, denotes on the vertical axis the Hamming distortion 
of the best
model for the noisy cross with respect to the noiseless cross
at the rate given on the horizontal axis.
The line hits almost zero distortion (Hamming distance 3)
at model cost bit rate about 250.  The best model of the noisy cross
at this rate is attached to this point. (The three wrong bits
are at the bottom left corner and upper right armpit.)
 This coincides with a sharp slowing
of the rate of decrease of the first graph. Subsequently, the second graph
rises again because the best model for the noisy cross starts to model
more noise. Thus, the second graph shows us the denoising of the noisy
cross, underfitting left of the point of contact with the horizontal axis,
and overfitting right of that point. This point of best denoising can 
also be deduced
from the first graph, where it is the point where 
the distortion-rate curve sharply
levels off. 
Since this point
has distortion of only $3$ to the
noiseless cross, the distortion-rate
function separates structure and noise very well in this
example. 

In the experiments in \cite{RV06} a specially written
block sorting compression algorithm with a
move-to-front scheme as described in \cite{BW94} was used. 
The algorithm is very
similar to a number of common general purpose compressors, such as bzip2
and zzip, but it is simpler and faster for small inputs; the source
code (in C) is available from the authors of \cite{RV06}.

\section{Algorithmic versus Probabilistic Rate-Distortion}\label{sect.algprobrd}

Theorem~\ref{thm.dresf} shows that
Shannon's rate-distortion function
$r^n(\delta)$ of \eqref{eq.rndelta}
for a random variable is pointwise related
to the expected value
of the rate-distortion functions $r_x(\delta)$ of the individual
string $x \in {\A}_n$ 
(outcomes of the random variable with the expectation taken
over the probabilities of the random variable). 
This result generalizes \cite{YS93,MK94,SE03}
to arbitrary computable sources.

Formally, probabilistic rate-distortion theory is treated in
Appendix~\ref{sect.ratedistortion}. 
Let ${\mathbf X}$ and ${\mathbf Y}$ be finite alphabets where we
take ${\mathbf X}=\{0,1\}$ for convenience.
We generalize the setting from i.i.d.
random variables to more general random variables.
Let $X_1, X_2, \ldots , X_n$ be a sequence
of, possibly dependent, random variables with values in ${\mathbf X}^n$
such that 
$p(x_1x_2\ldots x_n) = P(X_1=x_1, X_2=x_2, \ldots , X_n=x_n)$ 
is rational.  With $X=X_1, X_2, \ldots , X_n$ and
$x=x_1x_2 \ldots x_n$,
let $\K(X)$ denote
the Kolmogorov complexity of the set of pairs 
$(x,p(x))$ ordered lexicographic.
Let 
$E: {\mathbf X}^n \rightarrow {\mathbf Y}^n$ be a code.
Define the Shannon rate-distortion function by 
\begin{equation}\label{eq.rndelta}
r^n(\delta) = \min_E
\{ \log |E({\mathbf X}^n)| :
{\bf E} d(x,E(x)) \leq \delta \},
\end{equation}
the expectation ${\bf E}$
taken over the probability mass function $p$.

\begin{theorem}\label{thm.dresf}
Let 
$E_0$ be a many-to-one coding function 
defined by $E_0(x)=y$ with
$d(x,y) \leq \delta$ and
$r_x(\delta) = \K(y)$. 
Let $|x|=n$. Then,
\[
 {\bf E} r_x(\delta) - \Delta_1
\leq  r^n (\delta)
\leq \min \left\{{\bf E} r_x(\delta)+ \Delta_2 ,  
 \max_{x \in {\cal X}^n} r_x(\delta) \right\},
\]
with $\Delta_1 = O(\K(\delta,r^n,X,n))$,
$\Delta_2 = H(L)-H(S)$ with $S(y)= \sum \{p(x): E_0(x)=y\}$,
$L(y)$ is the uniform distribution over the $y$'s over $\mathbf{Y}^n$, and
the expectation ${\bf E}$ is taken over $p$.
\end{theorem}
Note that we have taken ${\cal X}= {\X}= \mathbf{X}^n$
and ${\Y}=\mathbf{Y}^n$.
The $\Delta_1$ quantity satisfies $\lim_{n \rightarrow \infty} \Delta_1 /n =0$.
The quantity $\Delta_2$ is small only in the case where we
have asymptotic equidistribution. This is the original setting of Shannon.
Though independence is not needed, for example ergodic stationarity guarantees
asymptotic equidistribution.

\appendix

\subsection{Shannon Rate Distortion}\label{sect.ratedistortion}
Classical rate-distortion theory
was initiated by Shannon in \cite{Sh48,Sh59}, and 
we briefly recall his approach.
Let 
${\mathbf X}$ and ${\mathbf Y}$ be finite alphabets.
A single-letter distortion measure is
a function $d$ that maps elements of
$\mathbf X \times \mathbf Y$ to the reals. Define the distortion between
word $x$ and $y$ of the same length $n$ over alphabets
${\mathbf X}$ and ${\mathbf Y}$, respectively, by
\[
d^n(x,y)= \frac{1}{n}\sum_{i=1}^n d(x_i,y_i).
\]
Let $X$ be a random variable with values in
${\mathbf X}$. Consider the random variable $X^n$ with values in ${\mathbf X}^n$,
that is, the sequence $X_1,\dots,X_n$ of $n$ independent
copies of $X$.
We want to encode words of length $n$ over ${\mathbf X}$ by words over ${\mathbf Y}$
so that the number
of all code words is small and
the expected distortion between outcomes of $X^n$ and their
codes is small.
The tradeoff between the expected
distortion
and the number of code words used is expressed
by the {\em rate-distortion} function
denoted by $r^n(\delta )$ as in \eqref{eq.rndelta}. It
maps every $\delta \in {\cal R}$
to the minimal natural number
$r$ (we call $r$ the \emph{rate})
having the following property:
There is an encoding function
$E:{\mathbf X}^n \rightarrow {\mathbf Y}^n$ with a range of cardinality at most $2^r$
such that
the expected distortion between the outcomes of $X^n$
and their corresponding codes is at most~$\delta$.

In \cite{Sh59} Shannon gave the following nonconstructive
asymptotic characterization of $r^n(\delta)$.
Let $Z$ be a random variable with values in ${\mathbf Y}$.
Let $H(Z)$, $H(Z \mid X)$ stand for the  Shannon entropy and conditional Shannon entropy,
respectively. Let $I(X;Z)=H(Z)-H(Z \mid X)$ denote the mutual information
in $X$ and $Z$, and  ${\bf E} d(X,Z)$ stand
for the expected value of  $d(x,z)$ with respect to
the joint probability $P(X=x, Z=z)$ of the random variables $X$ and $Z$. 
For a real $\delta$, let $R(\delta)$ denote
the minimal $I(X;Z)$ subject to ${\bf E} d(X,Z)\leq \delta$.
That such a minimum is attained for all $\delta$ can be shown
by compactness arguments.

\begin{theorem}\label{theo.shannon}
For every $n$ and $\delta$ we have
$r^n(\delta)\geq nR(\delta)$. Conversely,
for every $\delta$ and every positive $\epsilon$,
we have
$r^n(\delta+\epsilon)\leq n(R(\delta)+\epsilon)$
for all large enough $n$.
\end{theorem}

\subsection{Computability}\label{sect.computability}

In 1936 A.M. Turing \cite{Tu36} defined the hypothetical `Turing machine' 
whose computations are 
intended to give an operational and formal definition 
of the intuitive notion of computability in the discrete domain.
These Turing machines compute integer functions, 
the {\em computable} functions. By using pairs of integers for the 
arguments and values we can extend computable functions
to functions with rational arguments and/or values.
The notion of computability can be further
extended, see for example \cite{LiVi97}:
A 
function $f$ with rational arguments and real values is
{\em upper semicomputable}
if there is a computable
function  $\phi (x,k)$ with 
$x$ an rational number and $k$ a nonnegative integer
such that $\phi(x,k+1) \leq \phi(x,k)$ for every $k$ and
  $\lim_{k \rightarrow \infty} \phi (x,k)=f(x)$.
This means
  that $f$ can be computably approximated from above.
A function $f$ is
{\em lower semicomputable}
  if $-f$ is upper semicomputable.
  A function is called
{\em semicomputable}
  if it is either upper semicomputable or lower semicomputable or both.
If a function $f$ is both upper semicomputable and
lower semicomputable,
then $f$ is 
computable.
A countable set $S$ is {\em computably (or recursively) enumerable}
if there is a Turing machine $T$ that outputs all and only the elements of $S$
in some order and does not halt. A countable set $S$ is 
{\em decidable (or recursive)}
if there is a Turing machine $T$ that decides for every candidate $a$
whether $a \in S$ and halts. 

\begin{example}\rm
An example of a computable function is $f(n)$ defined as
the $n$th prime number;
an example of a function that is upper semicomputable
but not computable is the Kolmogorov complexity function $\K$ in
Appendix~\ref{sect.kolmcompl}. An example of a recursive set is the set
of prime numbers; an example of a recursively enumerable
set that is not recursive is $\{x \in {\cal N}: \K(x) < |x| \}$.
\end{example}

Let ${\cal X}=\{0,1\}^*$, and ${\Y}$ and the distortion measure $d$ 
be given.
Assume that ${\Y}$ is recursively (= computably) enumerable
and the set 
$\{\pair{x,y,\delta}\in\mathcal X\times\Y\times\Q: d(x,y)\le \delta\}$
is decidable.  
Then $r_x$ is upper semicomputable. Namely, to determine $r_x(\delta)$
proceed as follows.
Recall that $U$ is the reference universal Turing machine.
Run $U(p)$ for all $p$ 
dovetailed fashion (in stage $k$ of the overall computation
execute the $i$th computation step of the $(k-i)$th program).
Interleave this computation
with a process that recursively enumerates  ${\Y}$. 
Put all enumerated elements of ${\Y}$ in a set ${\cal W}$. 
Whenever $U(p)$ halts we put the output in a set ${\cal U}$.
After every step in the overall computation we determine the 
minimum length of a program $p$ such that $U(p) \in {\cal W} \bigcap {\cal U}$ 
and $d(x,U(p))\le\delta$. 
We call $p$ a \emph{candidate} program.
The minimal length of all candidate programs can only decrease
in time and eventually becomes equal to $r_x(\delta)$. Thus,
this process 
upper semicomputes $r_x (\delta)$. 

The function $g_x$ is also upper semicomputable. The proof is similar
to that used to prove the upper semicomputability of $r_x$.
It follows from \cite{VV02} that in general $d_x$,
and hence its `inverse' $r_x$ and by Lemma~\ref{lem.rg}
the function $g_x$, are not computable. 

Assume that the set $\Y$ is recursively enumerable and
the set
$\{\pair{x,y,\delta}\in\mathcal X\times\Y\times\Q: d(x,y)\le \delta\}$
is decidable. Assume that the resulting distortion family $\mathcal
A^{d,\Y}$
satisfies Property 2.
There is a relation between
destination words and distortion balls. This relation is as follows.

(i) Communicating a destination word $y$ for a source word $x$
 knowing a rational upper bound
$\delta$ for  the distortion $d(x,y)$
involved is the same as communicating a
distortion ball of radius $\delta$ containing $x$.

(ii) Given (a list of the elements of) a distortion ball $B$
we can upper semicompute
the least distortion $\delta$ such that $B=B(y,\delta)$ for some $y\in\Y$.

Ad (i). This implies that the function $\tilde r_x(\delta)$ defined
in \eqref{eq.tilde} differs from $r_x(\delta)$
by $O(\K(\delta)+\log |x|)$.
See the proof of Lemma~\ref{lem.rg}.

Ad (ii). Let
$B$ be a given ball. Recursively enumerating ${\Y}$ and 
the possible $\beta\in \Q$,
we find
for every newly enumerated element of $y \in {\Y}$
whether $B(y, \beta)=B$ (see the proof of  Lemma~\ref{lem.rg}
for an algortihm to find a list of elements of $B(y, \beta)$
given $y,\beta$). Put these $\beta$'s
in a set ${\cal W}$.
Consider the least element of ${\cal W}$ at every computation step.
This process upper semicomputes the
least distortion  $\delta$ corresponding to
the distortion ball $B$.

\subsection{Kolmogorov Complexity}\label{sect.kolmcompl}

For precise definitions, notation, and results see the text \cite{LiVi97}.
Informally, the Kolmogorov complexity, or algorithmic entropy, $\K(x)$ of a
string $x$ is the length (number of bits) of a shortest binary
program (string) to compute
$x$ on a fixed reference universal computer
(such as a particular universal Turing machine).
Intuitively, $\K(x)$ represents the minimal amount of information
required to generate $x$ by any effective process.
The conditional Kolmogorov complexity $\K(x \mid  y)$ of $x$ relative to
$y$ is defined similarly as the length of a shortest binary program
to compute $x$, if $y$ is furnished as an auxiliary input to the
computation.

Let $T_1 ,T_2 , \ldots$ be a standard enumeration
of all (and only) Turing machines with a binary input tape,
for example the lexicographic length-increasing ordered syntactic
Turing machine descriptions, \cite{LiVi97},
and let $\phi_1 , \phi_2 , \ldots$
be the enumeration of corresponding functions
that are computed by the respective Turing machines
($T_i$ computes $\phi_i$).
These functions are  the
{\em computable (or recursive)}
functions. 
For the development of the theory we
actually require
the Turing machines to use {\em auxiliary} (also
called {\em conditional})
information, by equipping the machines with a special
read-only auxiliary tape containing this information at the outset.
Let $\langle \cdot , \cdot \rangle$ be a computable one to one 
{\em pairing function}
on the natural numbers (equivalently, strings)
mapping $\{0,1\}^* \times \{0,1\}^* \rightarrow \{0,1\}^*$ with
$|\langle u,v \rangle| \leq |u|+|v| +O(\log (|u|))$. (We need the extra
$O(\log (|u|))$ bits to separate $u$ from $v$. 
For Kolmogorov complexity, it is essential that there 
exists a pairing function such that  
the length of $\langle u,v \rangle$ is equal to the sum of 
the lengths of $u,v$ plus a small value depending only on $|u|$.)
We denote the function computed by a Turing machine $T_i$ with $p$ as input
and $y$ as conditional information by
$\phi_i(p,y)$.

One of the main achievements of the theory of computation
is that the enumeration $T_1,T_2, \ldots$ contains
a machine, say $T_u$, that is computationally universal in that it can
simulate the computation of every machine in the enumeration when
provided with its index. It does so by computing a 
function $\phi_u$ such that 
   $\phi_u(\langle i, p\rangle,y)  = \phi_i (p,y)$
    for all $i,p,y$.
    We fix one such machine and designate it as the {\em reference universal
    Turing machine} or {\em reference Turing machine} for short.

\begin{definition}\label{def.KolmK}
    The {\em conditional Kolmogorov complexity} of $x$ given $y$ (as
auxiliary information) {\em with respect to Turing machine} $T_i$ is
                  \begin{equation}\label{eq.KC}
    \K_i(x \mid y) = \min_p \{|p|: \phi_i(p,y)=x \}.
                  \end{equation}
The {\em conditional Kolmogorov complexity} $\K(x \mid y)$ is defined
as the conditional Kolmogorov complexity 
$\K_u (x \mid y)$ with respect to the reference  Turing machine $T_u$ 
usually denoted by $U$.
The {\em unconditional} version is set to  $\K(x)=\K(x  \mid \epsilon)$.
\end{definition}  

Kolmogorov complexity $\K(x\mid y)$ has 
the following crucial property: 
$\K(x\mid y)\le \K_i(x \mid y)+c_i$ for 
all $i,x,y$, where $c_i$ depends only on
$i$ (asymptotically, the reference Turing machine is not worse
than any other machine).
Intuitively, $\K(x\mid y)$ represents the minimal amount of information
required to generate $x$ by any effective process from input $y$.
The functions $\K( \cdot)$ and $\K( \cdot \mid  \cdot)$,
though defined in terms of a
particular machine model, are machine-independent up to an additive
constant
 and acquire an asymptotically universal and absolute character
through Church's thesis, see for example \cite{LiVi97}, 
and from the ability of universal machines to
simulate one another and execute any effective process.
  The Kolmogorov complexity of an individual finite object was introduced by
Kolmogorov \cite{Ko65} as an absolute
and objective quantification of the amount of information in it.
The information theory of Shannon \cite{Sh48}, on the other hand,
deals with {\em average} information {\em to communicate}
objects produced by a {\em random source}.
 Since the former theory is much more precise, it is surprising that
analogs of theorems in information theory hold for
Kolmogorov complexity, be it in somewhat weaker form.
For example, let $X$ and $Y$ be random variables
with a joint distribution. Then,
$H(X,Y)\le H(X)+H(Y)$,
where $H(X)$ is the entropy of the marginal
distribution of $X$. 
Similarly, let $\K(x,y)$ denote $\K(\langle x,y \rangle)$ 
where $\langle \cdot,\cdot \rangle$
is a standard pairing 
function as defined previously and $x,y$ are strings.
Then we have 
$\K(x,y)\le \K(x)+\K(y)+O(\log \K(x))$. Indeed, there is a
Turing machine $T_i$ that provided with  $\langle p,q\rangle$ 
as an input computes $\langle U(p),U(q)\rangle$  
(where $U$ is the reference Turing machine). By construction of $T_i$, we have
$\K_i(x,y)\le \K(x)+\K(y)+O(\log \K(x))$, hence
$\K(x,y)\le \K(x)+\K(y)+O(\log \K(x))$.

Another interesting similarity is the following:
$I(X;Y)=H(Y)-H(Y \mid X)$
 is the (probabilistic)
{\em information in random variable $X$ about random variable $Y$}.
Here $H(Y \mid X)$ is the conditional entropy of $Y$
given $X$.
Since $I(X;Y)=I(Y;X)$ we call this symmetric quantity the {\em
mutual (probabilistic) information}. 
\begin{definition}
\label{def.mi}
\rm
The {\em (algorithmic)  information in $x$ about $y$} 
is $I(x:y)=\K(y)-\K(y\mid x)$,
where $x,y$
are finite objects like finite strings or finite sets of finite strings.
\end{definition}

It is  remarkable that also the algorithmic information
in one finite object about another one is symmetric: $I(x:y)=I(y:x)$ up to
an additive term logarithmic in $\K(x)+\K(y)$. This follows
immediately from the {\em symmetry of information} property
due to A.N. Kolmogorov and L.A. Levin: 
\begin{align}\label{eq.soi}
\K(x,y) & = \K(x)+\K(y \mid x) + O(\log (\K(x)+\K(y))) \\
& = \K(y)+\K(x \mid y)+O(\log (\K(x)+\K(y))) .
\nonumber
\end{align}

\subsection{Randomness Deficiency and Fitness}\label{sect.rd}
Randomness deficiency of an element $x$ of
a finite set $A$ according to Definition~\ref{def.rd} is
related with the fitness of $x \in A$ (identified with the fitness
of set $A$ as a model for $x$) in the sense of $x$ having most properties 
represented by the set $A$. Properties are identified with large
subsets of $A$ whose Kolmogorov complexity is small (the `simple'
subsets).
\begin{lemma}\label{lemma.property}
Let $\beta , \gamma$ be constants.
Assume that $P$ is a subset of $A$ with
$|P| \geq (1-2^{- \beta })|A|$  and
$\K(P\mid A)\leq \gamma$.
Then the randomness deficiency $\delta(x \mid A)$ of every
$x\in A \setminus P$ satisfies 
$\delta(x \mid A)> \beta-\gamma-O(\log \log |A|)$
\end{lemma}
\begin{proof} 
Since $\delta (x \mid A) = \log |A|-\K(x\mid A)$
and $\K(x\mid A) \leq \K(x\mid A,P)+\K(P\mid A) + O(\log \K(x\mid A,P))$,
while $\K(x\mid A,P) \leq - \beta + \log |A|+O(1)\le \log |A|+O(1)$,
we obtain
$\delta(x \mid A)> \beta-\gamma-O(\log \log |A|)$.
\end{proof}

The randomness deficiency measures our disbelief
that $x$ can be obtained
by random sampling in $A$ (where all elements of $A$ are
equiprobable). 
For every $A$, the randomness deficiency of almost all
elements of $A$ is small:
The number of $x\in A$ with $\delta(x \mid A)>\beta$ is fewer than
$|A|2^{-\beta}$. This can be seen as follows. 
The inequality $\delta(x \mid A)>\beta$ implies
$\K(x \mid A)<\log |A|-\beta$.
Since $1+2+2^2+\dots+2^{i-1}=2^i-1$, 
there are less than $2^{\log  |A|-\beta}$
programs of fewer than
$\log |A|-\beta$ bits. Therefore, 
the number of $x$'s satisfying
the inequality 
$\K(x\mid A)<\log |A|-\beta$ cannot be larger.
Thus, with high probability  
the randomness
deficiency of an element
randomly chosen in  $A$ is small.
On the other hand, if $\delta(x \mid A)$ is small,
then there is no way to refute the hypothesis
that $x$ was obtained
by random sampling from $A$: Every such
refutation is based on a simply described property
possessed by a majority of elements 
of $A$ but not by $x$. Here it is important that we consider
only simply described properties, since otherwise
we can refute the hypothesis by exhibiting the property
$P=A \setminus \{x\}$.

\subsection{Covering Coefficient for Hamming Distortion}\label{sect.exhamming}

The authors find it difficult to believe that the covering result
in the lemma below is new. But neither a literature search nor the
consulting of experts has turned up an appropriate reference.
\begin{lemma}\label{l2}
Consider the distortion family ${\cal H}_n$.
For all $0 \leq d\leq \delta\leq  \frac{1}{2}$ every Hamming ball of radius
$\delta$ in ${\cal H}_n$
can be covered by at most
$\alpha_n b(\delta)/b(d)$
Hamming balls of radius $d$ in ${\cal H}_n$,
where $\alpha_n $ is a  polynomial in $n$.
\end{lemma}

\begin{proof}
%
Fix a ball with center $y$ and radius $\delta = j/n \leq \frac{1}{2}$ where
$j$ is a natural number.
All the strings in the ball that are
at Hamming distance at most $d$ from $y$
can be covered by one ball
of radius $d$ with center $y$.
Thus it suffices,
for every $\Delta$ of the form $i/n$ with 
$i= 2,3, \ldots ,j$ 
(such
that $d<\Delta\leq \delta$), to cover
the set of all the strings at distance precisely $\Delta$ from $y$
by  $n^{c+1} b(\delta)/b(d)$ balls of radius $d$ 
for some fixed constant $c$.
Then the ball $B(y, \delta)$ is covered by at most 
$j n^{c+1} b(\delta)/b(d) \leq n^{c+2} b(\delta)/b(d)$ balls of 
radius $d$.

Fix
$\Delta$ and let the Hamming sphere $S$ denote the set of all 
strings at distance precisely 
$\Delta$ from $y$.
Let $f$ be the solution to the equation
$d+f(1-2d)=\Delta$ rounded to the closest rational of the form $i/n$.
Since $d<\Delta\leq  \delta\leq\frac{1}{2}$
this equation has a unique solution and
it lies in the closed real interval
$[0,1]$.
Consider a ball $B$ of radius $d$ with a random center $z$
at distance
$f$ from $y$. Assume that
all centers at distance $f$ from $y$ are chosen with equal probabilities
$1/s(f)$ where $s(f)$ is the number of points in a Hamming
sphere of radius $f$.
\begin{claim}\label{claim.prball}
Let $x$ be a particular string in $S$. Then
\[
\Pr( x \in B) \geq \frac{b(d)}{n^c b(\delta)}
\]
for some fixed positive constant $c$.
\end{claim}

\begin{proof}
Fix a string $z$ at distance  $f$ from $y$. We first claim
that the ball $B$ of radius $d$ with center
$z$ covers $b(d)/n^c$ strings in $S$.
Without loss of generality, 
 assume that the string $y$ consists of only  zeros
and string $z$ consists of $fn$ ones and $(1-f)n$ zeros.
Flip a set of $fd n$ ones
and a set  of
$(1-f)d n$ zeros in $z$ to obtain a string $u$.
The total number of flipped bits is equal to
$d n$ and therefore $u$ is at distance $d$ from
$z$. The number of ones in $u$ is
$fn-fd n+(1-f)d n=\Delta n$ and
therefore $u \in S$.
Different choices of the positions of the same numbers of flipped bits
result in different strings in
$S$. The number of ways to choose the flipped bits is equal to
$$
\binom{fn}{fd n}\binom{(1-f)n}{(1-f)d n}.
$$
By Stirling's formula, this is at least
$$
2^{fnh(d)+(1-f)nh(d)-O(\log n)}=
2^{nh(d)-O(\log n)}\ge
\frac{b(d)}{n^c},
$$
where the last inequality follows from \eqref{binom-entropy}.
Therefore a ball $B$ as above covers at least $b(d)/n^c$ strings
of $S$.
The probability
that a ball $B$, chosen uniformly at random as above,
covers a particular string $x\in S$ is the same for every such $x$
since they are in symmetric position.
The number of elements in a Hamming sphere 
is smaller than the cardinality of a Hamming ball of the same radius,
$|S| \leq b(\delta)$.
Hence with probability
$$
\frac{b(d)}{n^c |S|}\ge
\frac{ b(d)}{n^c b(\delta)} 
$$
a random ball $B$ covers a particular string $x$ in $S$.
\end{proof}

By Claim~\ref{claim.prball}, 
the probability that a random ball $B$ does not cover a particular
string $x \in S$ is at most $1-b(d)/(n^c b(\delta))$.
The probability that no ball out of $N$ randomly drawn such
balls $B$ covers 
a particular $x \in S$ (all balls are equiprobable) is at most 
\[
\left(1-\frac{ b(d)}{n^c b(\delta)}\right)^N
< e^{-N b(d)/(n^c  b(\delta))} .
\]
For $N = n^{c+1}  b(\delta)/ b(d)$,
the exponent of the 
right-hand side of the last inequality is  $-n$,
and the probability that $x$ is not covered is at most $e^{-n}$. 
This probability remains exponentially small even after
multiplying by $|S| \leq 2^n$, the number of different $x$'s in $S$.
Hence, with probability at least $1- (2/e)^n$ 
we have that $N$ random balls
of the given type cover all the strings in $S$. 
Therefore, there exists a deterministic selection of $N$
such balls that covers all the strings in $S$.
The lemma is proved.
(A more accurate calculation shows that
the lemma holds with $\alpha_n=O(n^4)$.)
\end{proof}

\begin{corollary}\label{cor.l2}
\rm
Since all strings of length $n$ are either in the Hamming ball
$B(00\ldots 0, \frac{1}{2})$ or in the Hamming ball
$B(11\ldots 1, \frac{1}{2})$ in ${\cal H}_n$,
the lemma implies that the set $\{0,1\}^n$
can be covered by at most
\[
N =  \frac{2\alpha_n  2^{n}}{b(d)}
\]
balls of radius $d$ for every $0 \leq d \leq \frac{1}{2}$.
(A similar, but direct, calculation lets us
replace the factor $2\alpha_n$ by $n$.)
\end{corollary}
%
%
%

\subsection{Proofs of the Theorems}
\label{sect.proofs}

\begin{proof}
{\em of Theorem}~\ref{theo.allshapesrd}.
(i)  Lemma~\ref{lem.shapesg} (assuming properties 1 through 4) 
implies that
the canonical structure function $g_x$ of every string $x$ of length
$n$ is close to some function in the family $G_n$. This can be seen
as follows. Fix $x$ and
construct $g$ inductively for $n, n-1, \ldots , 0$. Define
$g(n)=0$
and
$$
g(l-1)=\left\{\begin{array}{ll}
g(l)+1 & \text{if } g(l)<g_x(l-1),\\
g(l) & \text{otherwise.}
\end{array}\right.
$$
By construction this function belongs
to the family $G_n$.
Let us show that
$
g_x(l)=g(l)+O(\log n)$.
First, we prove that
\begin{equation}\label{eq.left}
g(l) \leq g_x(l)
\end{equation}
by induction on $l=n,n-1, \ldots , 0$.
For $l=n$ the inequality is straightforward, since
by definition $g(n)=0$.
Let $0\le l\leq n$. 
Assume that $g(i)\le g_x(i)$ for $i=n,n-1, \ldots , l$.
If $g(l) < g_x(l-1)$ then $g(l-1)= g(l)+1$ and therefore 
$g(l-1) \leq g_x(l-1)$. If $g(l) \geq g_x(l-1)$ then
$g(l-1) = g(l) \geq g_x(l-1)\ge g_x(l)\ge g(l)$ and hence 
$g(l-1) = g_x(l-1)$.

Second, we prove that
\[
g_x(l)\le g(l)+O(\log n)
\]
for every $l=0,1,\ldots, n$.
Fix an $l$ and consider the least
$m$ with $l \leq m \leq n$ such that $g_x(m)=g(m)$.
If there is no such $m$ we take $m=n$ and observe
that $g_x(n)=O(\log n)= g(n)+ O(\log n)$.
This way, $g_x(m)=g(m)+O(\log n)$ and for every $l<l'\le m$
we have $g(l'-1)<g_x(l'-1)$ due to inequality \eqref{eq.left}
and definition of $m$.
Then 
$g_x(l'-1)>g(l'-1)\ge g(l')$, since we know that $g$ is nonincreasing.
Then, by the definition of $g$ we have $g(l'-1)=g(l')+1$.  Thus
we have
$g(l)=g(m)+m-l$.
Hence,
$g_x(l)\le g_x(m)+m-l+O(\log n) = g(m)+m-l+O(\log n)=g(l)+O(\log n)$,
where the inequality follows from Lemma~\ref{lem.shapesg},
the first equality from the assumption that $g_x(m)=g(m)+O(\log n)$,
and the second equality from the previous sentence.

(ii)
In Theorem IV.4 
in \cite{VV02} we proved a similar statement
for the special distortion family ${\cal L}$
with an error term of $O(\log n)$.
However, for the special case ${\cal L}$
we can let $x$ be equal to the first $x$
satisfying the inequality
$g_x(l)\ge g(l)-O(\log n)$ for every $l$.
In the general case this does not work any more.
Here we construct $x$ together with sets
ensuring the inequalities
$g_x(l)\le g(l)+O(\sqrt{n\log n})$ for every $l=0,\dots,n$.

The construction is as follows.
Divide the segment $\{0,1,\dots,n\}$ into
$N=\sqrt{n/\log n}$ subsegments of length $\sqrt {n\log n}$ each.
Let
$l_0=n>l_1>\dots>l_N=0$ denote the end points of the
resulting subsegments.

To find the desired $x$, we
run the nonhalting algorithm below that takes
$n$ and ${\A}_n$ as input 
together with the values
of the function $g$ in the points $l_0,\dots,l_N$.
Let $\delta (n)$ be a computable integer valued 
function of $n$ of the order $\sqrt {n\log n}$
that will be specified later. 
\begin{definition}
\rm
Let $i=0,1,\dots,N$.
A set $F\in\A_n$ is called {\em $i$-forbidden}
if $|F|\le 2^{l_i}$ and 
$\K(F) < g(l_i)-\delta (n)$.
A set is called {\em forbidden} if 
it is $i$-forbidden for some $i=0,1,\dots,N$.
\end{definition}
We wish to find an $x$ that is outside all forbidden sets
(since this guarantees that $g_x(l_i)\ge g(l_i)-\delta (n)$ for every $i$).
Since $\K(\cdot)$ is upper semicomputable, moreover 
property 3 holds, and we are also given $n$ and $g(l_0),\dots,g(l_N)$,
we are able to find all forbidden sets using the following
subroutine.

\textbf{Subroutine $(n,{\A}_n, g(l_0),g(l_1), \ldots , g(l_n))$:}
\begin{quote} 
for every 
$F\in \A_n$ 
upper  semicompute
$\K(F)$; every time we find
$\K(F) < g(l_i)-\delta (n)$ 
and $|F|\le 2^{l_i}$ for some $i$ and $F$, then print $F$.
{\bf End of Subroutine}
\end{quote}

This subroutine prints all the forbidden sets in some order. Let 
$F_1,\dots,F_T$ be that order. Unfortunately 
we do not know when the subroutine will 
print the last forbidden set. In other words, we do not 
know the number $T$ of forbidden sets. To overcome this problem,
the algorithm will run the subroutine and every time a new  
forbidden set $F_t$ is printed, the algorithm will  
construct {\em candidate sets}
$B_0(t),\dots,B_N(t)\in\A_n$ satisfying $|B_i(t)|\le 2^{l_i}$ and  
$\K(B_i(t)) \le g(l_i)+\delta (n)$  
and the following condition
\begin{equation}\label{eq.capcup}
\bigcap_{j=0}^{N}B_j(t) \setminus \bigcup_{j=1}^{t}
F_j\ne \emptyset ,
\end{equation}
for every $t=0,\dots,T$.
For $t=T$ the set $\bigcup_{j=1}^{t}
F_j$ is the union of all forbidden sets, which guarantees the bounds
$g(l_i)-\delta (n)\le g_x(l_i)\le g(l_i)+\delta (n)$
for all $x$ in the set in the left hand side of \eqref{eq.capcup}. 
Then we will 
prove that these bounds imply that 
$g(l)-\delta (n)\le g_x(l)\le g(l)+\delta (n)$
for \emph{every} $l=0,\dots,n$.
Each time a new forbidden set 
appears (that is, for every $t=1,\dots,T$) 
we will need to update candidate sets so that \eqref{eq.capcup} remains 
true. To do that we will maintain a stronger 
condition than just non-emptiness of the left hand side of \eqref{eq.capcup}.
Namely, we will maintain the following invariant:
for every $i=0,1, \ldots,  N$, 
\begin{equation}\label{eq.invariant}
\left| \bigcap_{j=0}^{i} B_j(t) \setminus \bigcup_{j=1}^{t}
F_j \right| \geq
2^{l_i-i-1}\alpha_n^{-i}.
\end{equation}
Note that for $i=N$ inequality \eqref{eq.invariant} implies
\eqref{eq.capcup}.

{\bf Algorithm 
$(n,{\A}_n, g(l_0),g(l_1), \ldots , g(l_n))$:}
\begin{description}
\item 
%
{\bf Initialize.}
Recall that $l_0=n$.
Define the set $B_t(0)=\booln$ for every $t$.
This set is in ${\A}_n$ by property 1.

{\bf for } $i := 1, \ldots , N$ {\bf do}

Assume inductively that 
$|B_0(0) \bigcap B_1(0) \bigcap \cdots \bigcap B_{i-1} (0)| 
\geq 2^{l_{i-1}} \alpha_n^{-i+1}$, where $\alpha_n$ 
denotes a polynomial upper bound of the covering
coefficient of distortion family ${\A}_n$ existing by property 4. 
(The value $\alpha_n$ can be computed from $n$.)
Note that this inequality is satisfied
for $i=1$.
Construct $B_{i}(0)$ by
covering $B_{i-1}(0)$ by at most
$\alpha_n 2^{l_{i-1}-l_{i}}$ sets of cardinality at most
$2^{l_{i}}$
(this cover exists in ${\A}_n$ by property 4).
Trivially, this cover also covers
$B_0(0)\bigcap\dots\bigcap B_{i-1}(0)$.
The intersection of at least one of the covering
sets with $B_0(0)\bigcap\dots\bigcap B_{i-1}(0)$ has cardinality at least
$$
\frac{2^{l_{i-1}}\alpha_n^{-i+1}}{\alpha_n 2^{l_{i-1}-l_{i}}}=
2^{l_{i}}\alpha_n^{-i}.
$$
Let $B_{i}(0)$ by the first such covering set in a given standard order.
{\bf od}

Notice that after the Initialization the invariant~\eqref{eq.invariant}
is true for $t=0$, as $\bigcup_{j=1}^tF_j=\emptyset$.
For every $t=1,2,\dots$ perform the following steps 1 and 2
maintaining the 
invariant~\eqref{eq.invariant}: 

\item {\bf Step 1.}
Run the subroutine and wait until $t$th forbidden set $F_t$ is printed 
(if $t>T$ the algorithms waits forever and never
proceeds to Step 2). 

\item{\bf Step 2.}

{\bf Case 1.} For every $i = 0,1, \ldots , N$ 
we have 
\begin{equation}
\label{eq.inv}
\left|\bigcap_{j=0}^i B_j(t-1) \setminus \bigcup_{j=1}^t
F_j \right| \geq 2^{l_i-i-1}\alpha_n^{-i}.
\end{equation} 
Note the this inequality has one more 
forbidden set compared to the invariant~\eqref{eq.invariant} 
for $t-1$ (the argument in $B_j(t-1)$), and thus may be false. 
If that is the case, then 
we let $B_i(t)=B_i(t-1)$ for every 
$i=1, \ldots , N$ (this setting maintains invariant~\eqref{eq.invariant}). 

{\bf Case 2.} Assume that 
\eqref{eq.inv} is false 
for some index $i$.
In this case 
find the least such index (we will use later that \eqref{eq.inv} 
is true for all $i'<i$). 

We claim that $i>0$. That is,  
the inequality \eqref{eq.inv} is true for $i=0$.
In other words, the 
the cardinality of $F_1\bigcup \cdots \bigcup F_t$ is not
larger than half
of the cardinality of $B_0(t-1)=\booln$.
Indeed, for every fixed $i$ the total cardinality of all the sets
of simultaneously cardinality at most $2^{l_i}$ 
and Kolmogorov complexity less than $g(l_i)-\delta (n)$ does not exceed
$2^{g(l_i)-\delta (n)}2^{l_i}$. 
Therefore, the total number of elements in 
$\bigcup_{j=1}^t F_t$
is at most
$$
\sum_{i=0}^N2^{g(l_i)-\delta (n) +l_i}\le
(N+1)2^{g(\dmax)-\delta (n) +n}=
(N+1)2^{n- \delta (n) }\ll 2^{n-1}= \frac{1}{2}\left|\booln \right|,
$$ 
where the first inequality follows since the function $g(l)+l$
is monotonic nondecreasing, the first equality since 
$g(\dmax)=0$ by definition,
and the last inequality since we will set $\delta(n)$
at order of magnitude $\sqrt{n \log n}$.


First let $B_k(t)=B_k(t-1)$ for all $k<i$ (this
maintains invariant~\eqref{eq.invariant} for all $k<i$).
To define $B_i(t)$ find a covering
of $B_{i-1}(t)$ by at most
$\alpha_n 2^{l_{i-1}-l_i}$
sets in ${\A}_n$ of cardinality at most $2^{l_i}$.
Since~\eqref{eq.inv} 
is true for index $i-1$, we have
\begin{equation}\label{eq.inter}
\left| \bigcap_{j=0}^{i-1} B_j(t) \setminus 
\bigcup_{j=1}^t
F_t \right|
 \geq
2^{l_{i-1}-i}\alpha_n^{-i+1}.
\end{equation}
Thus 
the greatest cardinality of an intersection of the set in \eqref{eq.inter}
with a covering set is at least
$$
\frac{2^{l_{i-1}-i}\alpha_n^{-i+1}}{\alpha_n 2^{l_{i-1}-l_i}}
= 2^{l_i-i}\alpha_n^{-i}.
$$
Let $B_i(t)$ be
the first such covering set in standard order.
Note that $2^{l_i-i}\alpha_n^{-i}$ is at least
twice the
threshold required by invariant~\eqref{eq.invariant}. 
Use the same procedure to obtain successively $B_{i+1}(t),\dots,B_N(t)$.
\end{description}

{\bf End of Algorithm}

Although the algorithm does not halt,
at some unknown time  the last forbidden set $F_T$ is enumerated.
After this time the candidate sets are not changed anymore.
The invariant \eqref{eq.invariant} with $i=N$ shows that the cardinality 
of the set in the left hand side of \eqref{eq.capcup} is 
positive 
hence the set is not empty.

Next we show that $\K(B_i(t))\le g(l_i)+\delta(n)$
for every $i$ and every $t=1,\ldots,T$. We will see  
that to this end it suffices to upperbound
the number of changes of each candidate set. 

\begin{definition}
\rm
Let $m_i$ be the {\em number of changes of $B_i$}
defined by 
$m_i = |\{t: B_i(t) \neq B_i (t-1), \; 1 \leq t\le T \}|$ for
$0 \leq i \leq N$.
\end{definition}
\begin{claim}\label{claim.mi}
\rm
$m_i \leq 2^{g(l_i)+i}$ for $0 \leq i \leq N$. 
\end{claim}
\begin{proof}
The Claim is proved by induction on $i$. For 
$i=0$ the claim is true,
since $l_0 = n$ and $g(n)=0$ while $m_0=0$ by
initialization in the Algorithm ($B(0)$ never changes). 

($i > 0$): assume that the Claim 
is satisfied for every $j$ with $0 \leq j < i$.
We will prove that $m_i\le 2^{g(l_i)+i}$ by counting
separately the number of changes of $B_i$ of different types.

{\bf Change of type 1.} The set $B_i$ is changed when 
\eqref{eq.inv} 
is false for an index strictly
less than $i$. 
The number of these changes is at most 
\[
m_{i-1} \leq 2^{g(l_{i-1})+i-1} \leq 2^{g(l_{i})+i-1},
\]
where the first inequality follows from the inductive assumption,
and the second inequality by the property of $g$ that it
is nonincreasing.
Namely, since $l_{i-1} > l_i$  we have
$g(l_{i-1}) \leq g(l_i)$.

{\bf Change of type 2.}  The inequality \eqref{eq.invariant} 
is false for $i$ and is true for all smaller indexes.

{\bf Change of type 2a.}
After the last change of
$B_i$ at least one $j$-forbidden set for some $j<i$  
has been enumerated.
The number of changes of this type is at most the number of
$j$-forbidden sets for $j=0,\dots,i-1$. For every such $j$  
these forbidden sets have by definition Kolmogorov complexity less than
$g(l_{j}) - \delta (n)$. 
Since $l_j \ge l_i$ and $g$
is monotonic nonincreasing we have
$g(l_{j}) \leq g(l_{i})$. 
Because there are at most $N$ of these $j$'s,
the number of such forbidden sets is at most
$$N2^{g(l_i)-\delta(n)}\ll 2^{g(l_i)},$$
since we will later choose
$\delta(n)$ of order $\sqrt{n \log n}$, 

{\bf Change of type 2b.}
Finally, for every change of this type, between the last
change of
$B_i$ and the current one
no candidate sets with indexes less than
$i$ have been changed and no $j$-forbidden  sets
with $j<i$ have been enumerated.
Since after the last change of $B_i$ the cardinality of the set in the 
left-hand side of \eqref{eq.invariant} was at least 
$2^{l_i-i} \alpha_n^{-i}$, which is twice the threshold 
in the right-hand side
by the restoration of the invariant in the Algorithm Step 2, Case 2,
the following must hold.
The cardinality of 
$\bigcup_{j=1}^t F_j$ increased  by  at least
$2^{l_i-i-1}\alpha_n^{-i}$ since the last change of $B_i$,
and this must be due to enumerating
$j$-forbidden sets for $j=i,\dots,N$.
For every such $j$ 
every $j$-forbidden
set has cardinality at most $2^{l_j}$
and Kolmogorov complexity less than
$g(l_{j}) - \delta (n)$. 
Hence the total number of elements in all
$j$-forbidden sets is less than $2^{l_j}2^{g(l_{j}) - \delta (n)}$.
Since $j\geq i$ and hence $l_j \leq l_i$ while $g(l)+l$
is monotonic nondecreasing we have
$g(l_{j})+l_j \leq g(l_{i})+l_i$.
Because there are at most $N+1$ of these $j$'s,
the total number of elements in all those sets does not exceed
$M=(N+1)2^{g(l_i)-\delta (n)+l_i}$.
The number
of changes of this type is not more than the total number $M$
of elements involved divided by the increments of size
$2^{l_i-i-1}\alpha_n^{-i}$. Hence it is not more than
$$(N+1)2^{g(l_i)-\delta (n)}2^{i+1}\alpha_n^{i}.$$
Let
\begin{align}\label{eq.deltan}
&\delta (n) \geq \log ((N+1)2^{i+10}\alpha_n^{i})
\; \; {\rm and }
\\&\delta (n) = 
O (N\log(2\alpha_n))=O(\sqrt{n/\log n} \; \log(2\alpha_n))=
O (\sqrt{n\log n}),
\nonumber
\end{align}
where the last equality uses that $\alpha_n$ is polynomial
in $n$ by property 4. 
Then,
the number of changes of type 2b is much less than  $2^{g(l_i)}$.
 The value of $\delta(n)$ can be computed from $n$.

Summing the numbers of changes of types 1, 2a, and 2b we obtain
$m_i \leq 2^{g(l_i)+i}$, completing the induction.
\end{proof}
\begin{claim}\label{claim.gx}
\rm
Every $x$ in the nonempty set  \eqref{eq.capcup} satisfies
$|g_x(l_i) -  g(l_i)| \leq \delta (n)$
with $\delta (n) = O(\sqrt{n \log n})$
for $i=0,1, \ldots , N$.
\end{claim}
\begin{proof}
By construction $x$  is not an element of any forbidden set
in $\bigcup_{t=1}^T F_t$, and therefore
\[
g_x(l_i) \geq g(l_i) - \delta (n)
\]
for every $i=0,1, \ldots , N$.
By construction $|B_i(T)| \leq 2^{l_i}$, and
to finish the proof it remains to show that 
$\K(B_i (T))
\leq g(l_i)+\delta (n)$ so that 
$g_x(l_i) \leq g(l_i)+\delta(n)$, 
for $i=0,1, \ldots,  N$.
Fix $i$. 
The set $B_i(T)$ can be 
described by a constant length 
program, that is $O(1)$ bits,
that runs the Algorithm and uses the following
information:
\begin{itemize}
\item
A description of 
$i$ in $\log N\le\log n$ bits.
\item
A description of 
the distortion family $\A_n$ in $O(\log n)$ bits by property 3.
\item
The values of $g$ in the points $l_0,\dots,l_N$
in $N\log n=\sqrt{n\log n}$ bits.
\item
The description of $n$ in $O(\log n)$ bits.
\item
The total number $m_i$
of changes (Case 2 in the Algorithm)
to intermediate versions of $B_i$ in $\log m_i$ bits.
\end{itemize}
We count the number of bits in the description of 
$B_i(T)$. The description is effective and by Claim~\ref{claim.mi} with
$i \leq N = \sqrt{n/\log n}$ it
takes at most $g(l_i) + O(\sqrt{n \log n})$ bits. So this is an
upper bound on the Kolmogorov complexity $\K(B_i(T))$. 
Therefore, for some $\delta(n)$ satisfying \eqref{eq.deltan} we have
\[
g_x(l_i) \leq g(l_i)+ \delta(n),
\]
for every $i = 0,1, \ldots, N$. 
The claim follows from the first and the last displayed 
equation in the proof.
\end{proof}

Let us show that the statement 
of Claim~\ref{claim.gx} 
holds not only for the subsequence of values $l_0,l_1, \ldots , l_N$
but for every $l=0,1, \ldots , n$,

Let $l_i \leq l \leq l_{i-1}$.
Both functions $g(l),g_x(l)$ are nonincreasing so that
\begin{align*}
&g(l)\in[g(l_{i-1}),g(l_{i})],\\
&g_x(l)\in[g_x(l_{i-1}),g_x(l_{i})]
\subseteq[g(l_{i-1})-O(\sqrt{n\log n}),g(l_{i})+O(\sqrt{n\log n})].
\end{align*}
By the 
spacing of the sequence of $l_i$'s
the length of the segment
$[g(l_{i-1}),g(l_{i})]$ is at most
$$
g(l_{i})-g(l_{i-1})\le l_{i-1}-l_{i}
 = \sqrt{n\log n}.
$$
If there is an $x$ such that Claim~\ref{claim.gx} 
holds for every $l_i$ with $i=0, \ldots , N$, then 
it follows from the above that
$|g(l)-g_x(l)|\le\sqrt{n\log n}+O(\sqrt{n\log n})$ for every $l=0,1, \ldots, n$.
\end{proof}
\vspace{.2in}

\begin{proof}
{\em of Theorem}~\ref{th-shannon-analog}.
We start with Lemma~\ref{th5} stating a combinatorial fact 
that is interesting
in its own right, as explained further in Remark~\ref{rem.previously}.

\begin{lemma}\label{th5}
Let $n,m,k$ be natural numbers and
$x$ a string of length $n$. Let ${\BB}$ be a family
of subsets of $\{0,1\}^n$ and 
${\BB}(x) = \{B \in {\BB}: x \in B \}$.  If 
${\BB}(x)$ has at least $2^m$ elements (that is, sets) of
Kolmogorov complexity less than $k$, then
there is an element in ${\BB}(x)$ of Kolmogorov complexity
at most $k-m+O(\K(\BB)+\log n +\log k+\log m)$.
\end{lemma}

\begin{proof}
Consider a game between Alice and Bob. They alternate moves
starting with Alice's move.
A move of Alice consists in producing a
subset of $\booln$. A move of
Bob consists in marking some sets previously produced by
Alice (the number of marked sets can be 0).
Bob wins if after every one of his moves 
every $x\in\X$ that is covered by at least $2^m$
of Alice's sets
belongs to a marked set.
The length of a play is decided by Alice. She 
may stop the game after any of Bob's moves. However the 
total number of her moves (and hence Bob's moves) 
must be less than $2^k$. 
(It is easy to see that without loss of generality
we may assume that Alice makes exactly $2^k-1$ moves.)
Bob can easily win if he marks every set produced by Alice.
However, we want to minimize the total number of marked sets.

\begin{claim}\label{l53}
Bob has a winning strategy
that marks at most $O(2^{k-m}k^{2}n)$ sets.
\end{claim}

\begin{proof}
We present an explicit 
strategy for Bob, which consists in
in executing at every move $t=1,2, \ldots ,2^k -1$
the following algorithm for the sequence 
$A_1, A_2, \ldots , A_t$ which has been produced by Alice until then.

\begin{description}
\item
{\bf Step 1.} 
Let $2^j$ be the largest power 
of $2$ dividing $t$. 
Consider the last $2^j$ sets in the sequence 
$A_1, A_2, \ldots , A_t$ and call them
$D_1,\dots,D_{2^j}$.
\item
{\bf Step 2.}
Let $T$ be the set of $x$'s that occur in at least 
$2^{m}/k$ of the 
sets $D_1,\dots,D_{2^j}$. 
Let $D_p$ be a set such that $|D_p\bigcap T|$ is maximal.
Mark $D_p$ (if there is more than one then choose the one with $p$ least)
and remove all elements of  $D_p\bigcap T$ from $T$.
Call the resulting set $T_1$. 
Let $D_q$ be a set such that $|D_q\bigcap T_1|$ is maximal
(if there is more than one then choose the one with $q$ least).
After removing all elements of $D_q\bigcap T_1$ from $T_1$
we obtain a set $T_2$. Repeat the argument until 
we obtain $T_{e_j} = \emptyset$.
\end{description}

Firstly, for the $j$ above we have
$e_j \leq \lceil 2^{j-m}kn\ln2\rceil$.
This is proved as follows. We have 
$$
\sum_{i=1}^{2^j}|D_i\bigcap T|\ge|T|2^{m}/k,
$$
since every $x\in T$ is counted at least $2^{m}/k$ times in the
sum in the left hand side.
Thus there is a set in the list $D_1, \ldots , D_{2^j}$ 
such that the cardinality of its intersection
with $T$ 
is at least $2^{-j}$ times the right hand side.
 By the choice of $D_p$ it is such a set
and  we have $|D_p\bigcap T|\ge |T|2^{m-j}/k$.

The set $T$ has lost at least a $(2^{m-j}/k)$th fraction of its
elements, that is, $|T_1|\le |T|(1-2^{m-j}/k)$. 
Since $T_1 \subseteq T$, obviously every element of $T_1$ 
(still) occurs in at least 
$2^{m}/k$ of the sets $D_1,\dots,D_{2^j}$.
Thus we can repeat the argument and 
mark a set $D_q$ with $|D_q\bigcap T_1|\ge |T_1|2^{m-j}/k$. 
After removing all elements of $D_q\bigcap T_1$ from $T_1$
we obtain a set $T_2$ that is at most a $(1-2^{m-j}/k)$th fraction 
of $T_1$, that is, $|T_2|\le |T_1|(1-2^{m-j}/k)$. 

Recall that we repeat the procedure $e_j$ times where $e_j$
is the number of repetitions until  $T_{e_j} = \emptyset$.
It follows that $e_j \leq \lceil 2^{j-m}kn\ln2\rceil$
since
$$
|T|(1-2^{m-j}/k)^{2^{j-m}kn\ln2}<|T|e^{-n\ln2}=|T|2^{-n}\le1.
$$

Secondly, for every fixed $j=0,1, \ldots, k-1$ 
there are at most $2^{k-j}$ different $t$'s ($t=1,2, \ldots , 2^k-1$)
divisible by $2^j$ 
and the number $d_j = 2^{k-j}e_j$
of marked sets we need
to use for this $j$ satisfies 
$d_j \leq 2^{k-j} 2^{j-m} kn\ln2 = 2^{k-m} kn \ln2$.
For all $j=0,\dots,k-1$ together we use a total number of marked sets of
at most
\[
 \sum_{j=0}^{k-1} d_j \leq 2^{k-m} k^2 n\ln 2.
\]
In this way, 
after every move $t=1, 2,\ldots , 2^k-1 $ of Bob,
every $x$ occurring in
$2^m$ of Alice's sets belongs to a marked set of Bob.
This can be seen as follows.
Assume to the contrary, that there is an $x$
that occurs in $2^m$ of Alice's sets following move $t$ of Bob,
and $x$ belongs to no set marked by Bob in step $t$ or earlier.
Let $t= 2^{j_1} + 2^{j_2} + \cdots $ with $j_1>j_2>\cdots $
be the binary expansion of $t$. By Bob's strategy, 
the element $x$ occurs less than
$2^{m}/k$ times in the first segment of $2^{j_1}$ sets of Alice, 
less than $2^{m}/k$ times in the next segment of $2^{j_2}$ of Alice's 
sets, and so on.
Thus its total number of occurrences among the $t$ first sets of Alice is
strictly less than $k 2^m/k=2^m$.
The contradiction proves the claim.%
\end{proof}
Let us finish the proof of the Lemma~\ref{th5}.
Given the list of $\BB$,
recursively enumerate the sets in ${\BB}$ of Kolmogorov complexity 
less than $k$,
say $B_1, B_2, \ldots ,B_T$ with $T < 2^k$,
and consider this list as a particular sequence of
moves by Alice. 
Use Bob's 
strategy of Claim~\ref{l53} against Alice's 
sequence as above. 
Note that recursive enumeration of the sets in  ${\BB}$ 
of Kolmogorov complexity less than $k$ means that eventually all such
sets will be produced, although we do not know
when the last one is produced. This only means that the time between moves
is unknown, but the alternating moves between Alice and Bob are deterministic
and sequential.
According to Claim~\ref{l53}, Bob's strategy
marks at most
$O(2^{k-m}k^{2}n)$ sets.
These marked sets cover
every string occurring at least $2^m$ 
times in the sets $B_1, B_2, \ldots ,B_T$.
We do not know when the last set $B_T$ appears in this list,
but Bob's winning strategy of Claim~\ref{l53} ensures
that immediately after recursively enumerating $B_{i}$ 
$(i \leq T)$ in the list
every string that occurs in 
$2^m$ sets in the initial segment $B_1, B_2, \ldots B_{t}$
is covered by a marked set.
The Kolmogorov complexity $\K(B_i)$ of every marked set $B_i$
in the list $B_1, B_2, \ldots , B_T$ is upper bounded by
the logarithm 
of the number of
marked sets, that is
$k-m+O(\log k+\log n)$,
plus the description of ${\BB}$,
$k$, $m$, and $n$ including
separators in
$O(\K({\BB})+\log k+\log m+\log n)$ bits.
\end{proof}
We continue the proof of the theorem.
Let the distortion family ${\A}$ satisfy
properties 2 and 3.
Consider
the subfamily $\BB$ of $\A_n$ consisting of all sets $A$ with
$\wwh{\log A}=\wwh{\log B}$.
Let ${\BB}(x)$ be the family $\{B \in {\BB}: x \in B \}$ and
$N$ the number of sets in
${\BB}(x)$ of Kolmogorov complexity at most
$\K(B)$.

Given $x,\wwh{\log B},\A_n$ and $\K(B)$ 
we can generate all $A\in\BB(x)$ of Kolmogorov complexity
at most $\K(B)$.
Then we can describe $B$ by its index among the generated
sets. This shows that the description length 
$\K(B \mid x)\le \log N$
(ignoring an additive term of order $O(\log\K(B)+\log n)$ which suffices since
$\K(\wwh{\log B})$ and $\K(\A_n)$ are both $O(\log n)$).

Since $\K({\A}_n) = O(\log n)$ by property 3,
${\BB} \subseteq {\A}_n$ while every set $A \in {\BB}$ satisfies
$\lceil \log |A| \rceil = \lceil \log |B| \rceil \leq n$, we have
$\K({\BB}) = O(\log n)$. Let
$k=\K(B)+1$ and $m=\wh{\log N}$,
and ignore additive terms of order $O(\log k+\log m + \log n)$.
Applying  Lemma~\ref{th5} 
shows that there is a set  $A\in \BB(x)$
with $\K(A)\le k-m\le \K(B)-\K(B \mid x)=I(x:B)$ and therefore 
proves Theorem~\ref{th-shannon-analog}.
\end{proof}

\begin{remark}\label{rem.previously}
\rm
Previously an analog of Lemma~\ref{th5} was known in the case
when $\BB$ is the class of \emph{all} subsets $\booln$
of {\em fixed} cardinality  $2^l$.
For $l=0$ this is Exercise 4.3.8 (second edition) and 4.3.9
(third edition) of \cite{LiVi97}:
If a string $x$ has at least
$2^m$ descriptions of length at most $k$
($p$ is called a description of
$x$ if $U(p)=x$ where $U$ is
the reference Turing machine), then
$\K(x)\le k-m+O(\log k+\log m)$. Reference~\cite{VV02}
generalizes this to all $l> 0$:
If a string belongs to at least $2^m$
sets $B$ of cardinality $2^l$ and Kolmogorov complexity  $\K(B)\le k$,
then $x$ belongs to a set $A$ of cardinality $2^l$ and
Kolmogorov complexity 
$\K(A)\le k-m+O(\log m+\log k+\log l)$.
\end{remark}
\begin{remark}\label{rem.muchnik}
\rm
%
{\em Probabilistic proof of Claim~\ref{l53}.}
Consider a new game  that has the same rules and one additional
rule: Bob looses if he marks more than $2^{k-m+1}(n+1)\ln2$ sets.
We will prove that in this game Bob has a winning strategy.

Assume the contrary: Bob has no winning strategy. 
Since the number of moves in the game is finite (less than 
$2^k$), this implies that
Alice has a winning strategy.

Fix a winning strategy $S$ of Alice. To obtain a contradiction
we design a randomized strategy for Bob that beats Alice's
strategy $S$ with
positive probability. Bob's strategy is very simple:
mark every set produced by Alice with probability $p=2^{-m}(n+1)\ln2$.
\begin{claim}\label{claim.iii}
\rm
(i)
With probability more than $\frac{1}{2}$,
following every move of Bob every
element occurring in at least $2^m$ of Alice's sets is covered 
by a marked set of Bob.

(ii) With probability more than $\frac{1}{2}$, Bob marks
at most  $2^{k-m+1}(n+1)\ln2$ sets.
\end{claim}

\begin{proof}
(i) Fix $x$ and estimate
the probability that there is move of Bob following which $x$
belongs to $2^m$ of Alice's sets 
but belongs to no marked set of Bob.

Let $R_i$ be the event
``following a  move of Bob, string $x$
occurs at least in $i$ sets of Alice
but none of them is marked''.
Let us
prove by induction that
\[
\Pr [R_i]\le(1-p)^{i}.
\]
For $i=0$ the statement is trivial.
To prove the induction step we need to show that
$\Pr [R_{i+1}|R_i]\le 1-p$.

Let
$z=z_1,z_2,\dots,z_t$ be a sequence of decisions by Bob:
$z_j=1$ if Bob marks the $j$th set produced by Alice and
$z_j=0$ otherwise. Call $z$ \emph{bad} if
following Bob's $t$th move it happens
for the first time that $x$ belongs to $i$  sets produced by Alice
by move $t$ but none of them is  marked.
Then $R_i$ is the disjoint union of the events
``Bob has made the decisions $z$'' (denoted by $Q_z$) over all bad $z$.
Thus it is enough to prove that
\[
\Pr [R_{i+1} \mid Q_z]\le 1-p.
\]
Given that
Bob has made the decisions $z$, the event $R_{i+1}$
means that after those decisions the strategy $S$ will at some
time in the future produce the
$(i+1)$st set with member
$x$ but Bob will not mark it.
Bob's decision not to mark that set does not depend
on any previous decision and is made with probability $1-p$.
Hence
$$
\Pr [R_{i+1} \mid Q_z]=\Pr [\text{Alice produces 
the $(i+1)$st set with member }x \;  \mid  \;Q_z]
\cdot(1-p)
\le1-p.
$$
The induction step is proved.
Therefore,
$\Pr [R_{2^m}]\le (1-p)^{2^m}<e^{-p2^m}=2^{-n-1}$,
where the last equality follows by choice of $p$.

(ii) The expected number of marked sets is $p2^k$. Thus
the probability that it exceeds $p2^{k+1}$ is less than $\frac{1}{2}$.
\end{proof}

It follows from Claim~\ref{claim.iii} that there exists a strategy
by Bob that marks at most $2^{k-m+1}(n+1)\ln2$ sets out of Alice's
produced $2^k$ sets, and following every move of Bob every
element occurring in at least $2^m$ of Alice's sets is covered
by a marked set of Bob. Note that we have proved that
this strategy of Bob exists
but we have not constructed it.
Given $n$, $k$ and $m$, the number of games is finite, and  
a winning strategy for Bob can be found by brute force search.
%
%
%
%
\end{remark}

\vspace{.2in}
\begin{proof}{\em of Theorem~\ref{th45}}.
Let $B \subseteq\{0,1\}^n$ be a set containing string $x$. Define the
\emph{sufficiency deficiency of $x$ in $B$}
by
$$
\log|B|+\K(B)-\K(x).
$$
This is the number of extra bits incurred by the two-part code for $x$
using $B$ compared to the most optimal one-part code of $x$ using $\K(x)$ bits.
We relate this quantity with 
the randomness deficiency $\delta(x \mid B)=\log |B|-\K(x \mid B)$
 of $x$ in the set $B$.
The randomness deficiency is always less than the sufficiency
deficiency, and the
difference between them is equal to $\K(B \mid x)$:
\begin{equation}\label{eq76}
\log|B|+\K(B)-\K(x)-\delta(x \mid B)=\K(B \mid x),
\end{equation}
where the equality follows from the symmetry of 
information \eqref{eq.soi},
ignoring here and later in the proof additive terms of order
$O(\log\K(B)+\log n)$.

By Theorem~\ref{th-shannon-analog}, which assumes
that properties 2 and 3 hold for the distortion family
${\A}$, there is  $A\in\A(x)$
with $\wwh{\log|A|}=\wwh{\log|B|}$ and
$\K(A)\le \K(B)-\K(B \mid x)$.
Since $A_x$ is a set of minimal Kolmogorov complexity among
such $A$ we have
$\K(A_x)\le \K(B)-\K(B \mid x)$.
Therefore
\begin{align*}
\K(A_x)+\log|A_x|-\K(x)&\le\K(B)-\K(B \mid x)+\log|A_x|-\K(x)\\
&=
\K(B)-\K(B \mid x)+\log|B|-\K(x)=\delta(x \mid B),
\end{align*}
where the last equality is true by~\eqref{eq76}.
\end{proof}

\vspace{.2in}
\begin{proof}
{\em of  Theorem}~\ref{thm.dresf}.

{\em Left inequality.} 
Given $\delta$, $n$, $p$, and the (discrete) graph of $r^n$, we can compute an
optimal $E$ as in \eqref{eq.rndelta}  such that $r^n (\delta)
= \log |E(\mathbf{X}^n)|$. Retrieve $E(x)$ 
by its index of $r^n (\delta)$ bits in the set $E(\mathbf{X}^n)$.
Then,
\[
\K(E(x)) \leq r^n(\delta) + O(\K(\delta,r^n,X,n)).
\]
By definition, $r_x(\delta) \leq \K(E(x))$.
Taking the expectation of $r_x(\delta)$ over
$p$, we are done.

{\em Right inequality.}
Define a code $E_0$ such that
$\K(E_0(x)) = r_x(\delta)$
for every $x \in \mathbf{X}^n$.
Let $E_0(\mathbf{X}^n)$ be the range of $E_0$.
Although $E_0(\mathbf{X}^n)$ cannot be computed, it is finite, and trivially
\[
\log |E_0(\mathbf{X}^n)| \leq \max_{x \in \mathbf{X}^n} \K(E_0(x)).
\]
By definition $r^n(\delta) \leq \log |E_0(\mathbf{X}^n)|$, which yields
 $r^n (\delta)
\leq \max_{x \in \mathbf{X}^n} r_x(\delta)$.

The noiseless coding theorem, \cite{Sh48,LiVi97}, shows that
\[
\sum_{x \in \mathbf{X}^n} p(x)r_x(\delta) 
=  \sum_{y \in E_0(\mathbf{X}^n)} S(y) \K(y)
 \geq H(S),
\]
with $S$ the distribution defined in the statement of the theorem.
By definition, $r^n(\delta) \leq \log |\mathbf{Y}^n|$, which yields
$r^n(\delta) \leq H(L)$, with $L$ as in the statement of the theorem.
Together, we obtain
$r^n (\delta)
\leq {\bf E} r_x(\delta)+ \Delta_2$.
\end{proof}

\section*{Acknowledgements}
We thank Alexander K. Shen for helpful suggestions.
Andrei A. Muchnik gave the probabilistic proof
of Claim~\ref{l53} in Remark~\ref{rem.muchnik} after having seen
the deterministic proof. 
Such a probabilistic proof 
was independently proposed by Michal Kouck\'y.
We thank the referees for their constructive comments;
one referee pointed out that yet another example would be
the case of Euclidean balls with the usual Euclidean distance, where
the important Property 4 is proved in for example \cite{VG05}.
The work of N.K. Vereshchagin was done in part
while visiting CWI and was supported in part by the grant
09-01-00709 from Russian Federation
Basic Research Fund and by a visitors grant of NWO.
The work of P.M.B. Vit\'anyi was
supported in part by
the BSIK Project BRICKS
of the Dutch government and NWO, and by the
EU NoE PASCAL (Pattern Analysis, Statistical Modeling, 
and Computational Learning).

\end{document}